\documentclass{elsarticle}
\usepackage[utf8]{inputenc}
\usepackage{graphicx}
\usepackage{subfigure}
\usepackage{caption}
\usepackage{amsmath}
\usepackage{amsfonts}
\usepackage{bbm}
\usepackage{color}
\usepackage{hyperref}
\graphicspath{{./figures/}}

\def \br{\boldsymbol{r}}

\newcommand{\bsym}[1]{\boldsymbol{#1}}
\newcommand{\bbm}[1]{\mathbbm{#1}}

\newtheorem{algo}{Algorithm}

\begin{document}
\title{Machine learning materials physics: Surrogate optimization and multi-fidelity algorithms predict precipitate morphology in an alternative to phase field dynamics}
\author[umme]{Gregory H. Teichert}
\author[umme,umm,micde]{Krishna Garikipati\corref{mycorrespondingauthor}}
\ead{krishna@umich.edu}
\cortext[mycorrespondingauthor]{Corresponding Author}

\address[umme]{Mechanical Engineering Department, University of Michigan}
\address[umm]{Mathematics Department, University of Michigan}
\address[micde]{Michigan Institute for Computational Discovery \& Engineering, University of Michigan}

\begin{abstract}
    Machine learning has been effective at detecting patterns and predicting the response of systems that behave free of natural laws. Examples include learning crowd dynamics, recommender systems and autonomous mobility. There also have been applications to the search for new materials that draw upon big-data classification problems. However, when it comes to physical systems governed by conservation laws, the role of machine learning has been more limited. Here, we present our recent work in  exploring the role of machine learning methods in discovering, or aiding, the search for physics. Specifically, we focus on using machine learning algorithms to represent high-dimensional free energy surfaces with the goal of identifying precipitate morphologies in alloy systems. Traditionally, this problem has been approached by combining phase field models, which impose first-order dynamics, with elasticity, to traverse a free energy landscape in search of minima. Equilibrium precipitate morphologies occur at these minima. Here, we exploit the machine learning methods to represent high-dimensional data, combined with surrogate optimization, sensitivity analysis and multifidelity modelling as an alternate framework to explore phenomena controlled by energy extremization. This combination of data-driven methods offers an alternative to the imposition of first-order dynamics via phase field methods, and represents one aspect of applying machine learning to materials physics.
\end{abstract}


\maketitle

\section{Introduction}
Machine learning methods have been applied to a number of problems in materials physics, recently. These have included the search for compounds predicted by theory \cite{Hautier2010}, screening for new materials \cite{Meredig2014}, identifying stable compounds \cite{Raccuglia2016}, accelerated prediction of material properties \cite{Pilania2013}, the combination of data mining and quantum mechanics to predict crystal structures \cite{Fischer2006}, and many others. Studies in this set have used machine learning for classification by detecting patterns and making predictions in the face of complexity. A pair of recent studies have used Deep Neural Networks and Convolutional Neural Networks to recognize phase transitions as patterns in two dimensional Ising models \cite{Carrasquilla2017,Nieuwenburg2017}. Regression-based studies include the work of Kalidindi and co-workers, who have applied machine learning methods to extract properties based on materials microstructure \cite{Cecen2014,Steinmetz2016}. There also has been work on developing the plastic yield surface for a material using functional regression methods \cite{Versino2017}.  Here, we have studied whether machine learning, specifically Deep Neural Networks, in combination with other data-driven techniques, such as surrogate optimization, sensitivity analysis and multifidelity modelling, can be used to guide the prediction of precipitate shapes in alloys. Our study is framed by a well-defined physical principle: the Second Law of Thermodynamics in the form of minimization of free energy. 

A body of work has developed in the computational materials physics literature around the problem of precipitate growth in alloys. It includes a combination of methods: density functional theory for determination of free energies, elasticities, misfit strains and interfacial energies \cite{Thompson1999,Muller2001,Vaithyanathan2002,Ji2014,natarajanetal2016,Natarajan2017,Natarajan2017a}, continuum elasticity \cite{Su1996,Jou1997,Thompson1999,Muller2001,Vaithyanathan2002,Gao2012,Liu2013,Liu2017,Ji2014} and phase field dynamics for following the growth of precipitates \cite{Su1996,Jou1997,Leo1998,Muller2001,Hu2001,Zhu2002,Vaithyanathan2004,Gao2012,Liu2013,Ji2014,Liu2017}. Recently, these methods have been used to study the shapes of precipitates in magnesium-rare earth alloys \cite{Vaithyanathan2004,Gao2012,Liu2013,Ji2014,Liu2017}. The free energy parameterization in these works is in terms of chemical, elastic and interfacial contributions. Since elastic wave propagation is many orders of magnitude faster than diffusive transport in solids, the elasticity problem is considered at equilibrium for any state of the slower-evolving chemistry. Time dependence comes in the form of phase field dynamics of some conserved or non-conserved order parameters. If mass transport is important to the problem description, a conserved order parameter is defined, which is governed by classical transport-reaction equations, or by some version of the Cahn-Hilliard \cite{CahnHilliard1958} equation. A non-conserved parameter enters the description if the identities of the precipitate and matrix phases are delineated separately from the composition. The Allen-Cahn equation \cite{Allen1979} describes the corresponding dynamics. 

Whether conserved or non-conserved, the dynamics drive the system toward a free energy minimum, thus respecting the Second Law. Given that the system evolves over a complex, possibly high-dimensional, free energy landscape, we are prompted in this work to ask whether the equilibrium (minimum energy) states can be detected directly: by constructing the landscape ``on-the-fly'', and following it to minima. This approach, in principle, could have many advantages. The foremost is that while the dynamics must evolve in a serial fashion, following ``the arrow of time'', the construction of the free energy landscape is an embarrassingly parallel task: A large number of states, chosen either at random or by a sampling procedure, can be computed simultaneously to guide the search for a minimum free energy state. From the available states at any stage of this approach, the minima can be found by any of a number of procedures. Sorting is one approach if the states are directly used. However, using the states to define a representation for the free energy landscape make more possibilities accessible: Methods of sensitivity analysis and multi-fidelity modelling can be invoked to guide the search for minima through the possibly high-dimensional space of parameters. While phase field approaches could occasionally lead to trapped states in local minima, or to very slowly evolving dynamics, knowledge of the surrounding landscape could present opportunities for acceleration. Finally, it also is commonly observed that when sharp transitions occur between states of the system, the dynamics become stiff, leading to numerical difficulties with stability and therefore convergence. On the other hand, algorithms that sample states on the free energy landscape need not be restricted by high gradients with respect to states, making it potentially easier to escape past sharp transitions.

The preceding discussion leaves open the question of how the free energy landscape may be represented. For the problem of free energy minimizing precipitate shapes, functional representations would have natural parametrizations in terms of geometrical variables that describe the precipitate and its orientation, as well as the elasticities of the precipitate and matrix, and the interfacial energies. The space could vary between less than a dozen parameters if only geometry and average compositions matter, and reach as high as several dozen if elasticities and interfacial energies are included.

While it can be challenging to fit a function with even as few as three parameters, machine learning algorithms for regression may hold some potential in this regard. Given only a moderate degree of smoothness in the free energy, Neural Networks (due to their uniform approximation properties) could be effective, and are extendable to  higher dimensions without difficulty. Neural network-based approaches have encountered success at representing rapidly varying, high-dimensional data by increasing the number of hyper-parameters: hidden layers, thus becoming Deep Neural Networks (DNNs), and by increasing the number of units in any layer. Training a DNN is also a straightforward task, whether with \emph{de novo} code, or as is now the standard, working with the suite of open software available. In this work, we adopt DNNs to represent the high-dimensional free energy landscape. 

This is the task we set for the machine learning models: To represent the complexity that resides in the physics (phase segregation driven by elasticity, chemistry and interface effects) as a free energy landscape to guide the identification of equilibrium precipitate shapes. We note that the physics of energy minimization offers some novel approaches to training and expanding the data set, which can be cast as surrogate-based optimization methods following, for instance, the work of Vu et al. \cite{Vu2017}. 
A key ingredient of our approach is that direct numerical simulation (DNS) using nonlinear elasticity generates the data set for training the machine learning models. In turn the machine learning model, using the principle of energy minimization, populates the space with the parametric values at which further computations are to be carried out for precipitate free energy. Thus high performance computational physics and machine learning operate in a coupled fashion in our algorithms, each driving and being driven by the other to construct and explore the free energy landscape toward minima. This interaction can be further informed by an understanding of the influence of each parameter being explored, as provided by a sensitivity analysis. We refer to this approach as DNS-ML for direct numerical simulation-machine learning. Our method leverages a novel heterogeneous computing architecture for data-driven computational physics.

The number of DNS runs required to generate sufficient data for training the DNN can be significant. To limit the computation time required, techniques in multifidelity modelling can be applied \cite{Forrester2007,Kim2007,March2012,Bonfiglio2018}. Low-fidelity yet inexpensive data can provide insight to overall trends of the phenomena being modelled, while high-fidelity data is only required to provide a correction to the low-fidelity model. We take advantage of such techniques via Knowledge-Based Neural Networks (KBNNs). Thus, our surrogate model optimization routine also uses multifidelity optimization.

An outline of this paper is as follows: Section \ref{sec:phase field} describes the phase field model that we implemented as a baseline method for precipitate shape prediction. The machine learning approach is described in Section \ref{sec:DNS-ML}, with the algorithm and computing architecture being presented in Section \ref{sec:algo}. The phase field and DNS-ML predictions are compared in Section \ref{sec:results}. Section \ref{sec:conclusions} reviews the major conclusions to be drawn from this work.

\section{Phase field model}
\label{sec:phase field}
\begin{figure}[tb]
        \centering
\begin{minipage}[t]{0.7\textwidth}
        \centering
        \includegraphics[width=0.7\textwidth]{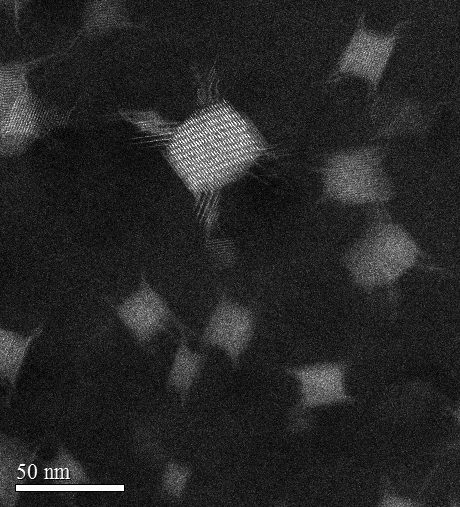}
        	\captionof{subfigure}{Incident beam parallel to $[0001]_\mathrm{Mg}$}
\end{minipage}
\begin{minipage}[t]{0.7\textwidth}
        \centering
	\includegraphics[width=0.9\textwidth]{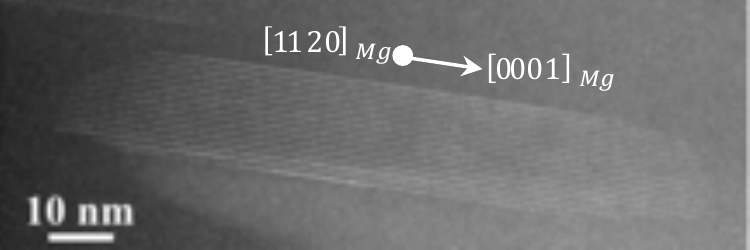}
	\captionof{subfigure}{Incident beam parallel to $[1120]_\mathrm{Mg}$}
\end{minipage}
	\caption{HAADF-STEM images of Mg–Y $\beta'$ precipitates (images by Ellen Solomon).}
	\label{fig:MgY_STEM}
\end{figure}

The aging process for a single $\beta'$ precipitate in Mg--2.8 atomic \% Y at 200$^\circ$ C (see Figure \ref{fig:MgY_STEM}) was simulated using a phase field model for the chemistry, coupled with finite strain elasticity. The crystal structure was described by an order parameter, $\eta$, where $\eta = 0$ corresponds to $\alpha$-Mg, and $\eta = 1$ to the ordered $\beta'$ precipitate. Values of $\eta$ between 0 and 1 define the diffuse interface. The composition of Y is described by a composition field variable, $c$.  The Kim-Kim-Suzuki (KKS) phase field model was used, in which the two phases in the diffuse interface are considered to have the same chemical potential rather than the same composition \cite{Kim1999}. The form of the equations used here closely follows that of Ji et al. \cite{Ji2014}. 

The reference and current placements of the precipitate-matrix system are denoted by $\Omega_0 \subset \mathbb{R}^3$ and $\Omega \subset \mathbb{R}^3$, respectively. The reference and current positions of a material point are $\bsym{X}\subset \mathbb{R}^3$ and $\bsym{x} = \bsym{\varphi}(\bsym{X}) \subset \mathbb{R}^3$, respectively. The displacement vector field, $\bsym{u} = \bsym{\varphi}-\bsym{X}$ is the primal variable for elasticity. It leads to the deformation gradient tensor, $\bsym{F} = \bsym{1} + \partial\bsym{u}/\partial\bsym{X} \in \mathbb{GL}^3$.

In the absence of boundary traction, the Gibbs free energy is defined by the following integral, defined on the reference configuration, $\Omega_0$, with :
\begin{equation}
\Pi[c,\eta,\bsym{u}] = \int_{\Omega_0} \left(f(c,\eta) + f_\text{grad}(\nabla\eta) + \psi(\bsym{F}^\text{e}(\eta,\bsym{F}),\eta)\right)\mathrm{d}V
\label{eqn:pf_energy}
\end{equation}
where $f(c,\eta)$ is the local chemical free energy density, $f_{grad}(c,\eta)$ is the gradient energy term, and $\psi(\bsym{F}^\text{e}(\eta,\bsym{F}),\eta)$ is the strain energy density, all defined per unit volume in $\Omega_0$. Note that, following convention, we define $c(\bsym{x}\circ\bsym{\varphi},t)$, $\eta(\bsym{x}\circ\bsym{\varphi},t)$ and $\bsym{u}(\bsym{X})$.

\subsection{Local free energy}
The local chemical free energy density includes the bulk chemical free energy and the Landau free energy densities describing the  structural change in the alloy. The bulk chemical free energy density is written in terms of contributions from the $\alpha$ and $\beta^\prime$ phases and a regularized Heaviside function $h(\eta)$, where $h(0) = 1$, $h(1) = 1$, and $h'(0) = h'(1) = 0$. The Landau free energy density has wells at $\eta = 0$ and $\eta = 1$.
\begin{align}
f(c,\eta) &= f^\alpha(c^\alpha)\left(1-h(\eta)\right)
+f^{\beta'}(c^{\beta'})h(\eta)
+\omega f_\text{Land}(\eta)\\
h(\eta) &= 3\eta^2 - 2\eta^3\\
f_\text{Land}(\eta) &= \eta^2-2\eta^3+\eta^4 \label{eqn:Landau}
\end{align}

\begin{figure}[tb]
\begin{minipage}[t]{0.5\textwidth}
        \centering
	\includegraphics[width=0.9\textwidth]{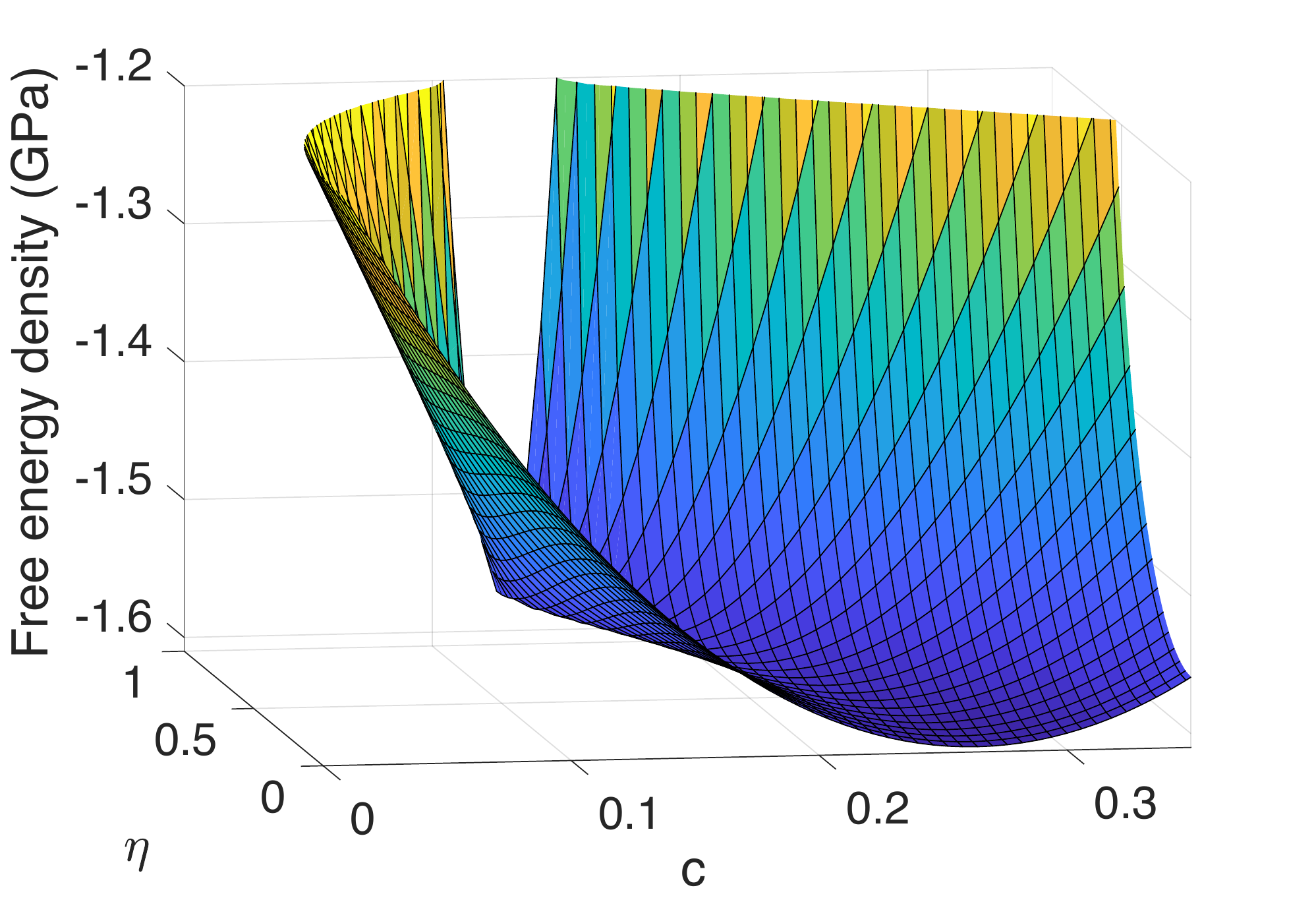}
\end{minipage}%
\begin{minipage}[t]{0.5\textwidth}
        \centering
	\includegraphics[width=0.8\textwidth]{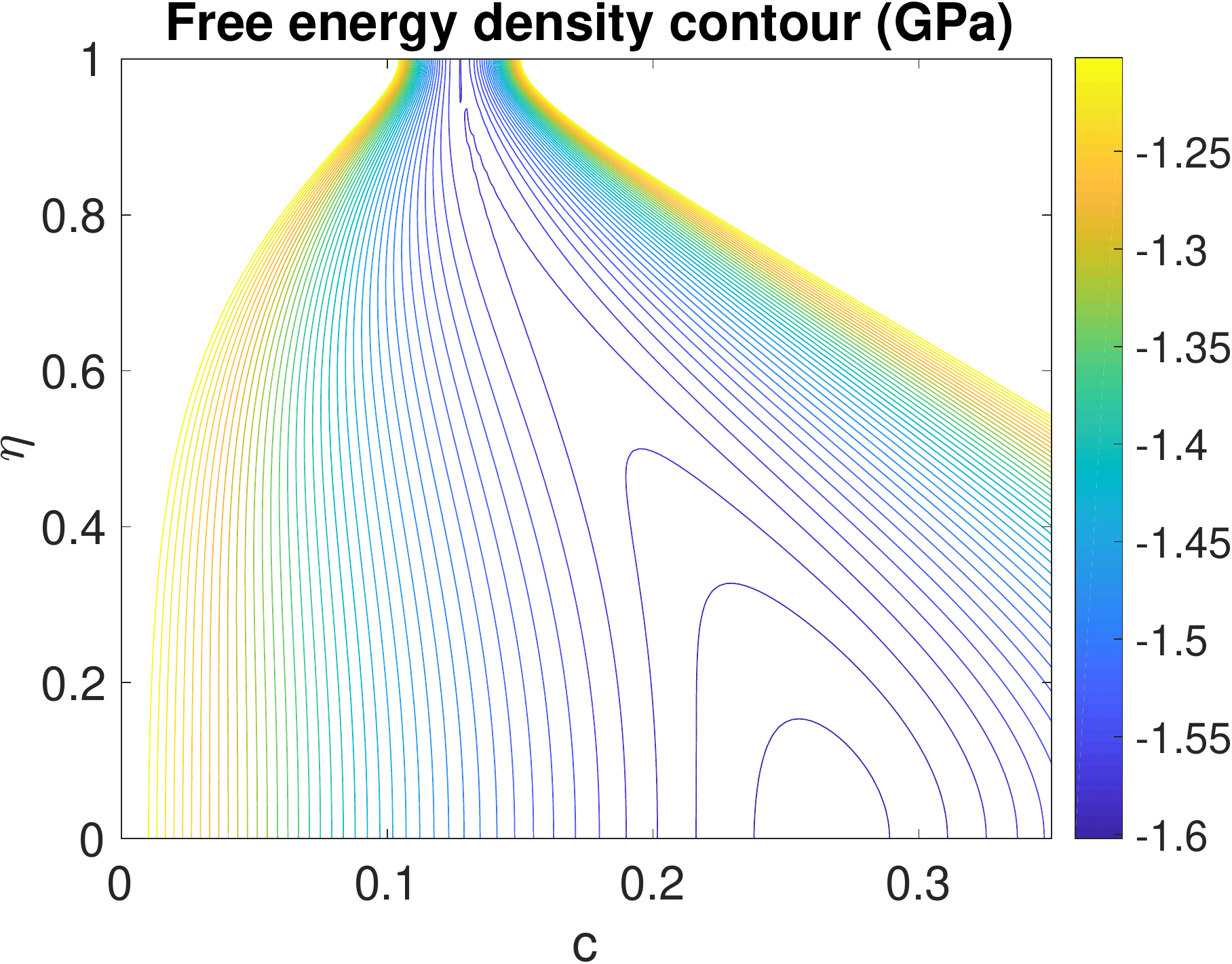}
\end{minipage}
    \caption{Surface \textcolor{black}{and contour plots} of the local free energy density used for the Mg-Y system in this study.}
	\label{fig:localFreeEnergy}
\end{figure}

The full form of the chemical free energy density of the Mg-Y solid solution is given by the following function \cite{Guo2007,Liu2013}:
\begin{equation}
\begin{split}
f^\alpha(c^\alpha) &= (1-c^\alpha)G_\mathrm{Mg} + c^\alpha G_\mathrm{Y} + c^\alpha(1-c^\alpha)(L_0 + L_1(1-2c^\alpha))\\
&\phantom{=} + RT((1-c^\alpha)\log(1-c^\alpha) + c^\alpha\log(c^\alpha))
\end{split}
\end{equation}
where the values for the coefficients at $473$K C are given in Table \ref{tab:chem}. The chemical free energy of the $\beta'$ precipitate is written such that a common tangent exists at $f^\alpha(0.01)$ and $f^{\beta'}(0.125)$ \cite{Liu2013}. The functions $f^\alpha(c^\alpha)$ and $f^{\beta'}(c^{\beta'})$ can be approximated as quadratic functions of the following form:
\begin{align}
    f^\alpha(c^\alpha) &\approx A^\alpha(c^\alpha - c^\alpha_0)^2 + B^\alpha \label{eqn:f_alpha}\\
    f^{\beta'}(c^{\beta'}) &\approx A^{\beta'}(c^{\beta'} - c^{\beta'}_0)^2 + B^{\beta'} \label{eqn:f_beta}
\end{align}
where the parameters are given in Table \ref{tab:chem2}. The resulting local free energy density is shown in Figure \ref{fig:localFreeEnergy}.

\begin{table}[tb]
    \centering
    \caption{Coefficients in the chemical free energy density for the Mg-Y solid solution at $473$K  (kJ mol$^{-1}$) \cite{Guo2007}.}
    \begin{tabular}{c | c }
    \hline
    $G_\mathrm{Mg}$ & -16.564\\
    $G_\mathrm{Y}$ & -21.561\\
    $L_0$ & -20.016\\
    $L_1$ & -2.836
    \end{tabular}
    \label{tab:chem}
\end{table}

\begin{table}[tb]
    \centering
    \caption{Parameters in the quadratic chemical free energy density descriptions.}
    \begin{tabular}{c | c c}
    \hline
    $A^\alpha$ & 6.2999 &GPa\\
    $B^\alpha$ & -1.6062 &GPa\\
    $c^\alpha_0$ & 0.2635\\
    $A^{\beta'}$ & 704.23 &GPa\\
    $B^{\beta'}$ & -1.5725 &GPa\\
    $c^{\beta'}_0$ & 0.1273
    \end{tabular}
    \label{tab:chem2}
\end{table}

The following constraint equations act to maintain the equal chemical potential condition and define the values $c^\alpha$ and $c^{\beta'}$:
\begin{align}
c &= (1-h(\eta))c^\alpha + h(\eta)c^{\beta'} \label{eqn:c}\\
\frac{\partial f}{\partial c} &= \frac{\partial f^\alpha}{\partial c^\alpha} = \frac{\partial f^{\beta'}}{\partial c^{\beta'}} \label{eqn:muEqual}
\end{align}
Eqs. (\ref{eqn:f_alpha}) and (\ref{eqn:f_beta}) can be used with Eqs. (\ref{eqn:c}) and (\ref{eqn:muEqual}) to find the following expressions for $c^\alpha$ and $c^{\beta'}$:
\begin{align}
c^\alpha &= \frac{A^{\beta'}\left[c - c^{\beta'}_0h(\eta)\right] + A^\alpha c^\alpha_0h(\eta)}{A^\alpha h(\eta) + A^{\beta'}(1 - h(\eta))} \label{eqn:calpha}\\
c^{\beta'} &= \frac{A^\alpha\left[c - c^\alpha_0(1-h(\eta))\right] + A^{\beta'} c^{\beta'}_0(1-h(\eta))}{A^\alpha h(\eta) + A^{\beta'}(1 - h(\eta))}  \label{eqn:cbeta}
\end{align}

\subsection{Gradient energy}
\label{sec:gradientenergy}
An anisotropic gradient energy term is used, with the second order tensor $\bsym{\kappa}$ being related to the anisotropic interfacial energy:
\begin{align}
f_\text{grad}(\nabla\eta) &= \frac{1}{2}\nabla\eta\cdot\bsym{\kappa}\nabla\eta
\label{eqn:grad_energy}
\end{align}
where $\nabla$ is the gradient operator in $\Omega_0$. 
The components of $\bsym{\kappa}$ are related to the barrier height $\omega$, the interface thickness, and the interfacial energy tensor $\bsym{\gamma}$ using the equilibrium solution for the one-dimensional problem and neglecting elasticity. The matrix of components of the tensor $\bsym{\gamma}$ is diagonal when $\bsym{\gamma}$ is written with respect to the Euclidean basis vectors $\bsym{e}_1,\bsym{e}_2$ and $\bsym{e}_3$ that coincide with the $[100], [010]$ and $[001]$ directions. The corresponding interfacial energies are the $\gamma_{11}, \gamma_{22}, \gamma_{33}$ components. Representing the thickness of the interface perpendicular to the $\bsym{e}_i$ direction as $(2\lambda_i)$, the following relations are used \cite{Kim1999}:
\begin{align}
\gamma_{ii} &= \frac{\sqrt{\kappa_{ii}\omega}}{3\sqrt{2}} &\text{(no sum over $i$)} \label{eqn:kappa}\\
2\lambda_i &= 2.2\sqrt{\frac{2\kappa_{ii}}{\omega}} &\text{(no sum over $i$)} \label{eqn:lambda}
\end{align}
Orientation variants can be considered by applying the appropriate rotation tensor to $\bsym{\gamma}$, introducing off-diagonal terms in $\bsym{\gamma}$ and $\bsym{\kappa}$. The interfacial energies for $\beta'$ precipitates in Mg-Y were calculated and reported by Liu et al \cite{Liu2013}. These values are shown in Table \ref{tab:interfacialEnergy}.
This study used the following values for $\bsym{\kappa}$ and $\omega$:
\begin{align}
\bsym{\kappa} = 
\begin{bmatrix}
0.1413 & 0 & 0\\
0 & 0.002993 & 0\\
0 & 0 & 0.1197
\end{bmatrix}\, \textcolor{black}{\mathrm{J/m},\,\omega = 0.115896\, \mathrm{J/m^3}}
\end{align}

\begin{table}[tb]
    \centering
    \caption{Interfacial energy between a $\beta'$ Mg-Y precipitate and the Mg matrix \cite{Liu2013}.}
    \begin{tabular}{c | c }
    Crystallographic & Interfacial\\
    plane & energy (J/m$^2$)\\
    \hline
    (100) & $\gamma_{11} = 0.03016$\\
    (010) & $\gamma_{22} = 0.00439$\\
    (001) & $\gamma_{33} = 0.02776$
    \end{tabular}
    \label{tab:interfacialEnergy}
\end{table}

\subsection{Strain energy}
The strain energy density function is the St.\ Venant-Kirchhoff model. The elasticity constants are modeled as being dependent on the order parameter to represent the difference in elasticity between the two phases. The total strain energy of the precipitate-matrix system is driven by a strain mismatch between the crystal structures of the matrix phase, $\alpha$, and precipitate phase, $\beta^\prime$. The stress-free transformation tensor of the $\beta'$ precipitate, $\bsym{F}_{\beta^\prime}$ (see Table \ref{tab:eigenstrain}), and the order parameter are used to determine the misfit strain, represented by $\bsym{F}^\lambda$. A multiplicative decomposition of the total deformation gradient into the parts due to elasticity and the misfit strain is used:
\begin{subequations}
\begin{align}
\psi(\bsym{F}^\text{e}(\eta,\bsym{F}),\eta) &= \frac{1}{2}\bsym{E}^\text{e}:(\bbm{C}^\alpha(1-h(\eta) + \bbm{C}^{\beta'}h(\eta)):\bsym{E}^\text{e}\label{eqn:SVK}\\
\bsym{E}^\text{e} &= \frac{1}{2}\left({\bsym{F}^\text{e}}^\mathsf{T}\bsym{F}^\text{e} - \bbm{1}\right)\\
\bsym{F}^\text{e}(\eta,\bsym{F}) &= \bsym{F}{\bsym{F}^\lambda}^{-1}(\eta)\label{eqn:FeFlam}\\
\bsym{F}^\lambda(\eta) &= \bbm{1}(1-h(\eta)) + \bsym{F}_{\beta^\prime}h(\eta)
\end{align}
\end{subequations}
The elasticity constants used for the Mg matrix were calculated and reported by Ji and co-workers \cite{jietal2014}. The elasticity constants for the matrix correspond with experimental data, although no such experimental data is available for the precipitate material. The elasticity constants used for the precipitate were calculated using density functional theory (see Table \ref{tab:elasticity}).

\begin{table}[tb]
    \centering
    \caption{Elasticity constants used for the Mg matrix \cite{jietal2014} and the $\beta'$ precipitate (calculated by Anirudh Natarajan, unpublished data) (GPa).}
    \begin{tabular}{c | c c }
     & Mg & $\beta'$\\
    \hline
    $\bbm{C}_{1111}$ & 62.6 & 78.8\\
    $\bbm{C}_{2222}$ & 62.6 & 62.9\\
    $\bbm{C}_{3333}$ & 64.9 & 65.6\\
    $\bbm{C}_{1122}$ & 26.0 & 24.6\\
    $\bbm{C}_{2233}$ & 20.9 & 19.9\\
    $\bbm{C}_{3311}$ & 20.9 & 23.1\\
    $\bbm{C}_{1212}$ & 18.3 & 11.9\\
    $\bbm{C}_{2323}$ & 13.3 & 11.6\\
    $\bbm{C}_{3131}$ & 13.3 & 8.46
    \end{tabular}
    \label{tab:elasticity}
\end{table}

\subsection{Initial and boundary value problems of phase field transport coupled with elasticity}
The phase field dynamics following the KKS model are modeled with the diffusion and Allen-Cahn equations, of the following forms \cite{Allen1979}:
\begin{align}
\frac{\partial c}{\partial t} &= -\nabla\cdot\bsym{J}\\
\frac{\partial \eta}{\partial t} &= -L\mu_\eta
\end{align}
where the flux is defined by $\bsym{J} := -M\nabla\mu_c$, $M$ is the mobility, and $L$ is the kinetic coefficient. The chemical potentials $\mu_c = \delta \Pi/\delta c$ and $\mu_\eta = \delta \Pi/\delta \eta$ are found using standard variational methods, giving the following expressions when assuming $\nabla\eta\cdot\bsym{\kappa}\bsym{n} = 0$ on $\partial \Omega$ (resulting from requiring equilibrium with respect to $\eta$ at the boundary, $\partial\Omega_0$):
\begin{align}
\mu_c &= \frac{\partial f^\alpha}{\partial c}\left(1-h(\eta)\right)+\frac{\partial f^{\beta'}}{\partial c}h(\eta)\\
\mu_\eta &= \left[f^{\beta'} - f^\alpha - \mu_c(c^{\beta'} - c^\alpha)\right]\frac{\partial h}{\partial \eta} - \nabla\cdot\bsym{\kappa}\nabla\eta + \omega\frac{\partial f_\text{Land}}{\partial \eta} + \frac{\partial \psi}{\partial \eta}
\end{align}
where
\begin{align}
\frac{\partial \psi}{\partial \eta} &= \left(\frac{1}{2}\bsym{E}:(\bbm{C}^{\beta'} - \bbm{C}^\alpha):\bsym{E} - 
\bsym{P}:\left(\bsym{F}^\text{e}(\bsym{F}^{\beta'} -\bbm{1})\textcolor{black}{\bsym{F}^{\lambda^{-1}}}\right)\right)\frac{\partial h}{\partial \eta}\\
\bsym{P} &= \frac{\partial\psi}{\partial \bsym{F}^\text{e}}
\end{align}

We now turn to the weak form of the phase field equations, with zero flux on all boundaries. We take $M$ and $L$ to be uniform and constant, and we define $\bar{\mu}_\eta := \mu_\eta + \nabla\cdot(\bsym{\kappa}\nabla\eta)$. The reference configuration and its boundary are denoted by $\Omega_0$ and $\partial\Omega_0$, respectively.
We seek solutions $c,\eta \in \mathcal{V} = \{w\in \mathcal{H}^1(\Omega_0)\}$ such that, for all weighting functions $w\in\mathcal{V}$,
\begin{align}
\int_{\Omega_0}\left(w\frac{\partial c}{\partial t} + \nabla w\cdot \left(M\nabla\mu_c\right)\right)\,\mathrm{d}V &= 0\\
\int_{\Omega_0}\left(w\left(\frac{\partial \eta}{\partial t} + L\bar{\mu}_\eta\right) + \nabla w\cdot \left(L\bsym{\kappa}\nabla \eta\right)\right)\,\mathrm{d}V &= 0
\end{align}
The initial conditions define a spherical precipitate of radius $r_0$, precipitate composition $c_{\mathrm{p}_0}$, and matrix composition $c_{\mathrm{m}_0}$ with a smoothed interface of width $\delta$:
\begin{align}
    c_0 &= 
    \begin{cases}
    c_{\mathrm{p}_0} & ||\bsym{X}|| < r_0-\frac{\delta}{2}\\
    0.5(c_{\mathrm{m}_0} - c_{\mathrm{p}_0})(||\bsym{X}|| - r_0+\frac{\delta}{2}) + c_{\mathrm{p}_0} & r_0-\frac{\delta}{2} \leq ||\bsym{X}|| < r_0+\frac{\delta}{2}\\
    c_{\mathrm{m}_0} & ||\bsym{X}|| \geq r_0+\frac{\delta}{2}
    \end{cases}\\
    \eta_0 &= 
    \begin{cases}
    1 & ||\bsym{X}|| < r_0-\frac{\delta}{2}\\
    -0.5(||\bsym{X}|| - r_0+\frac{\delta}{2}) + 1 & r_0-\frac{\delta}{2} \leq ||\bsym{X}|| < r_0+\frac{\delta}{2}\\
    0 & ||\bsym{X}|| \geq r_0+\frac{\delta}{2}
    \end{cases}\
\end{align}

The equilibrium conditions for mechanics are found by setting the first variation of the free energy functional with respect to the displacement equal to zero. Normal displacements are constrained to vanish at the boundary. Otherwise, zero traction boundary conditions are specified. The disjoint sets defining the Dirichlet and Neumann boundaries are denoted by $\partial\Omega_0^{u_i}$ and $\partial\Omega_0^{T_i}$, respectively, where $\partial\Omega_0^{u_i}\cup\partial\Omega_0^{T_i} = \partial\Omega_0$ for $i = 1,\ldots,3$. We seek a solution $\bsym{u}$ with $u_i \in \mathcal{S} = \{u_i\in\mathcal{H}^1(\Omega_0)|\bsym{u}\cdot\bsym{N}=0 \text{ on }\partial\Omega_0^{u_i}\}$ such that, for all weighting functions $\bsym{w}$ with $w_i\in\mathcal{W}=\{w_i\in\mathcal{H}^1(\Omega_0)|w_i=0 \text{ on }\partial\Omega_0^{u_i}\}$,
\begin{equation}
\int_{\Omega_0} \frac{\partial\bsym{w}}{\partial\bsym{X}}:\left(\bsym{P}{\bsym{F}^\lambda}^{-\mathsf{T}}\right)\,\mathrm{d}V = 0
\label{eqn:elastweakform}
\end{equation}
The corresponding strong form is the following:
\begin{align}
\mathrm{Div}\left(\bsym{P}{\bsym{F}^\lambda}^{-\mathsf{T}}\right) &= \bsym{0} \text{ in } \Omega_0\\
\left(\bsym{P}{\bsym{F}^\lambda}^{-\mathsf{T}}\right)\bsym{N} &= \bsym{0} \text{ on } \partial\Omega_{0}^{T_i}\\
\bsym{u}\cdot\bsym{N}&=0 \text{ on }\partial\Omega_0^{u_i}
\end{align}

\subsection{Numerical results}
The coupled phase field and non-linear elastic equilibrium equations were solved by the finite element method.The code\footnote{The code is available upon request and will soon be released as an application in the {\tt mechanoChemFEM} library developed by the authors and co-workers [\url{github.com/mechanoChem/mechanoChemFEM}]. {\tt mechanoChemFEM} is based on the {\tt deal.II} library.} was implemented in C++ using the {\tt deal.II} library \cite{bangerthetal2016}. The backward Euler time-stepping scheme was used for the phase field dynamics. The phase field simulations were run in parallel on Intel Xeon E5-2680 v3 processors on the XSEDE Comet HPC cluster \cite{Towns2014}. The phase field results are compared with results using a machine learning algorithm in Section \ref{sec:results}.

\subsubsection{Single precipitate}
\begin{figure}[tb]
        \centering
        \includegraphics[width=0.45\textwidth]{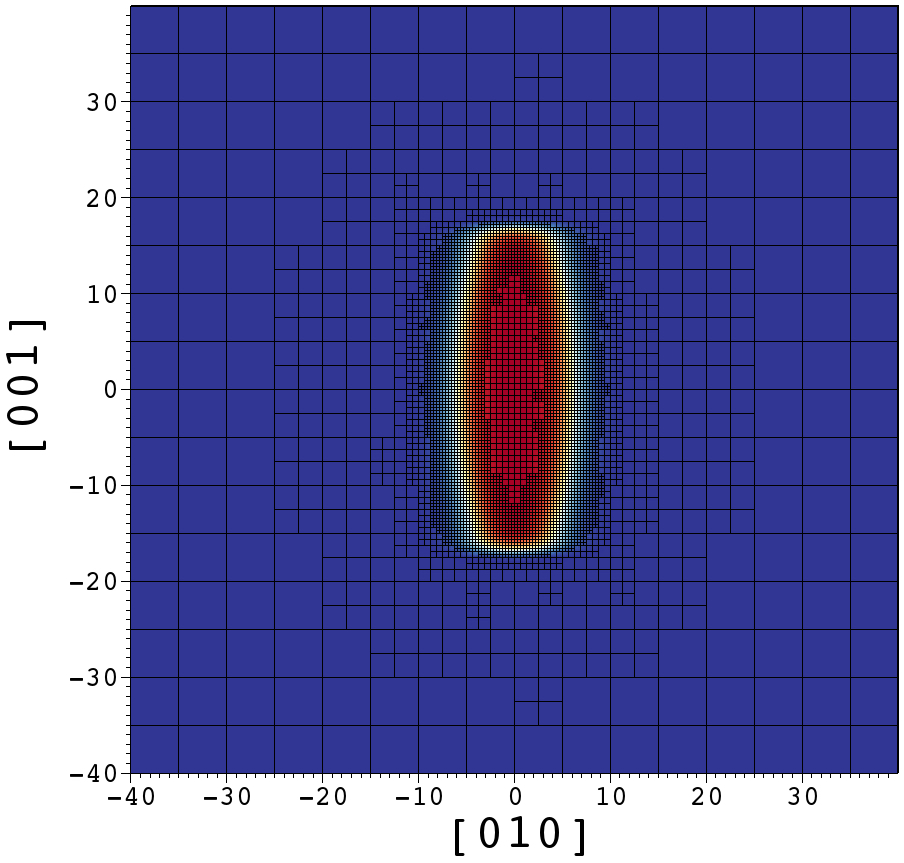}
        	\caption{2D slice of the phase field simulation showing the finite element mesh.}
	\label{fig:pf_mesh}
\end{figure}

A single precipitate was simulated in a Mg matrix domain of dimensions $80 \times 80 \times 110$ nm$^3$. The precipitate was initialized as a sphere with a volume of 6,000 nm$^3$, with $c_{\mathrm{p}_0} = 0.125$ and $c_{\mathrm{m}_0} = 0.02716$ to give an average Y composition of $c_\mathrm{avg} = 0.028$. Finite element meshes with up to 6M degrees of freedom were used (see Figure \ref{fig:pf_mesh} for an example of the finite element meshes used). The code was run in parallel on 240 physical cores over ten compute nodes.

\begin{figure}[tb]
        \centering
\begin{minipage}[t]{0.24\textwidth}
        \centering
	\includegraphics[width=0.95\textwidth]{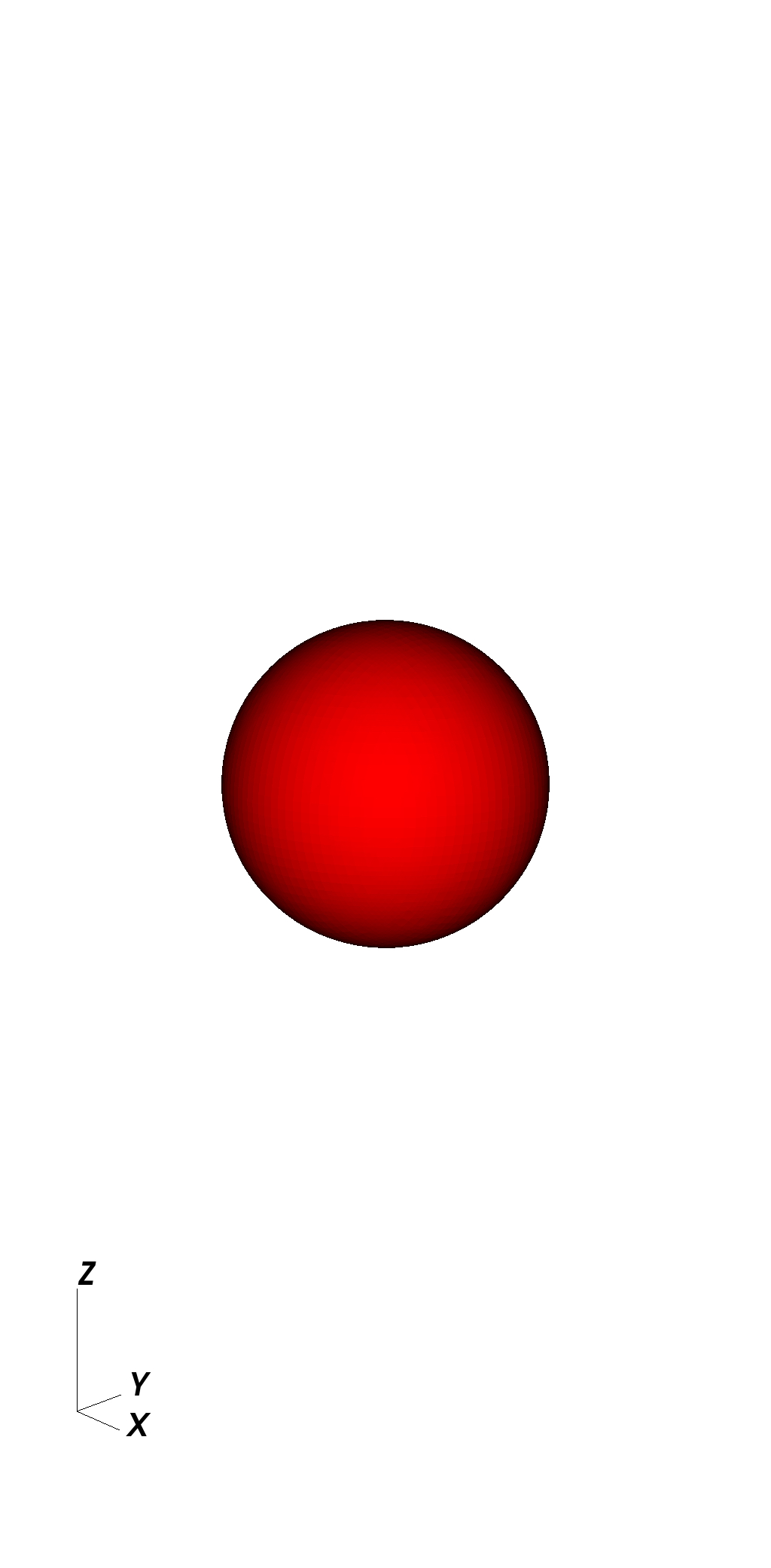}
	\captionof{subfigure}{Initial shape}
\end{minipage}
\begin{minipage}[t]{0.24\textwidth}
        \centering
	\includegraphics[width=0.95\textwidth]{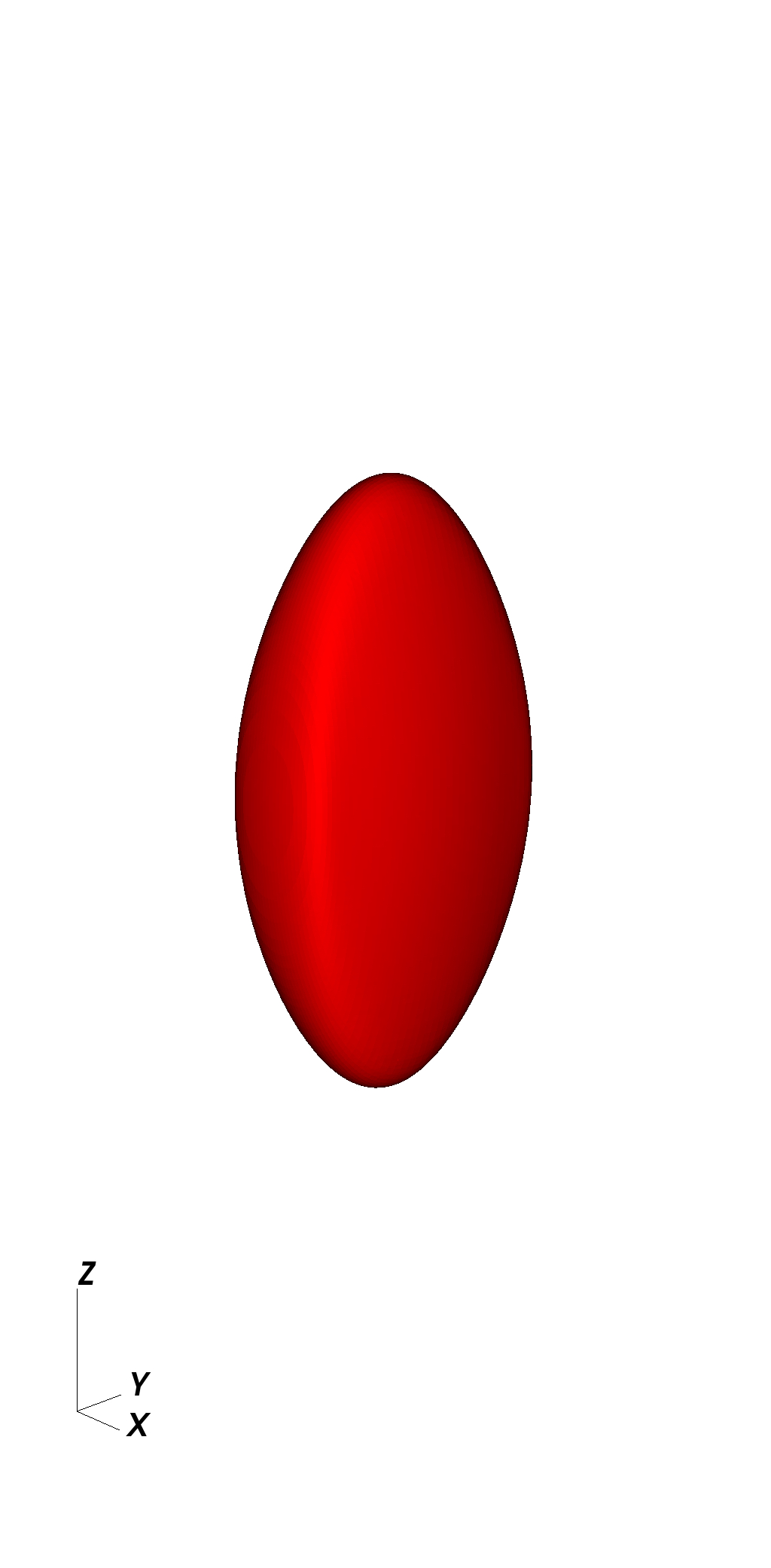}
	\captionof{subfigure}{Time step 200}
\end{minipage}
\begin{minipage}[t]{0.24\textwidth}
        \centering
	\includegraphics[width=0.95\textwidth]{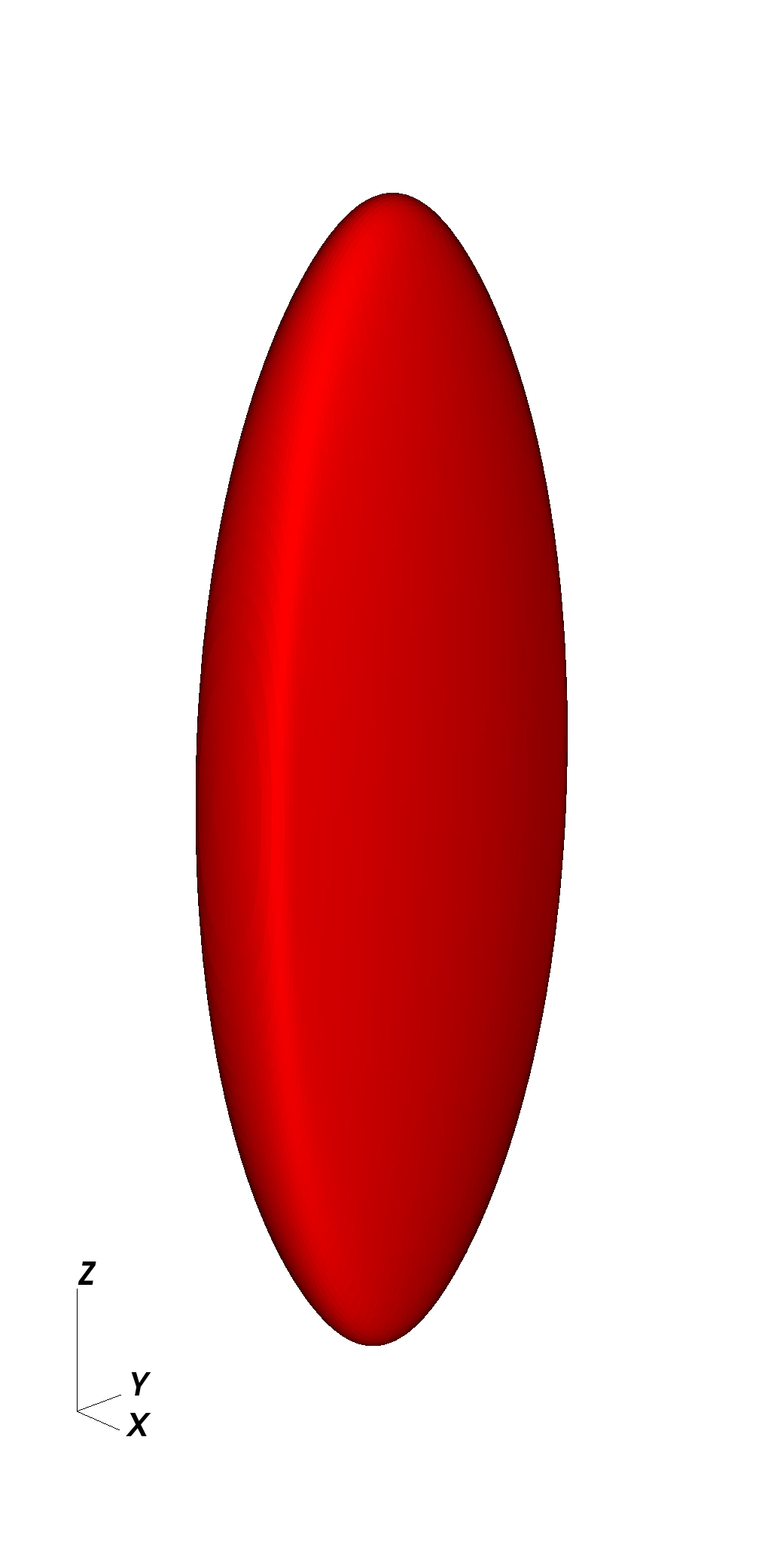}
	\captionof{subfigure}{Time step 400}
\end{minipage}
\begin{minipage}[t]{0.24\textwidth}
        \centering
	\includegraphics[width=0.95\textwidth]{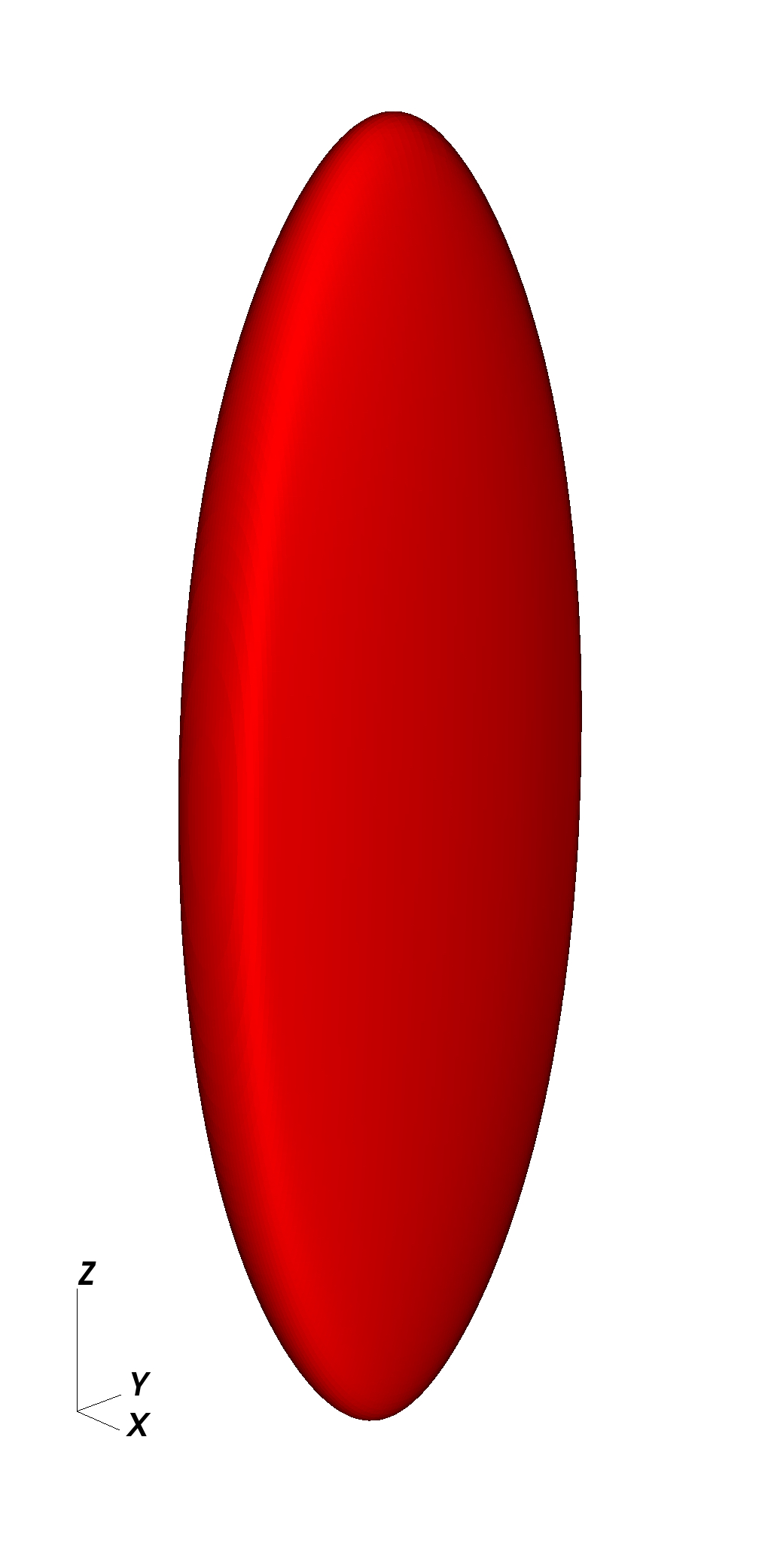}
	\captionof{subfigure}{Time step 600}
\end{minipage}
        \caption{Growth and evolution of the single precipitate over time as modeled by the phase field method.}
	\label{fig:PF_evolution}
\end{figure}

\begin{figure}[tb]
        \centering
        \includegraphics[width=0.55\textwidth]{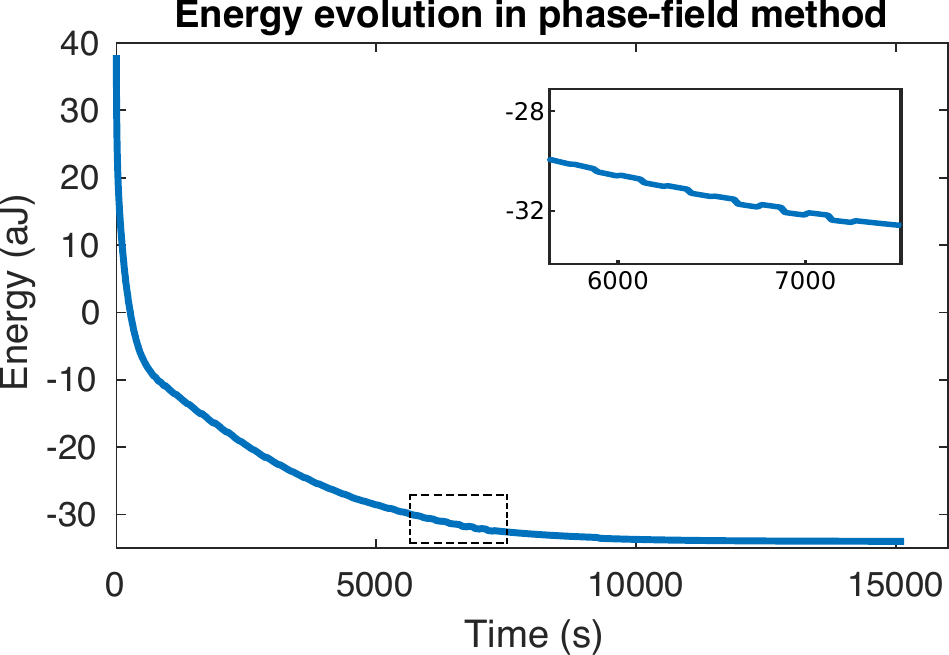}
        	\caption{Evolution of the total  free energy in the phase field method over simulated time. Slight fluctuations exist due to small, spurious changes in the global mass caused by adaptive mesh refinement and coarsening.}
	\label{fig:pf_energy}
\end{figure}

A near-equilibrium phase field solution was reached by 600 time steps. The evolution of the phase field solution over time appears in Figure \ref{fig:PF_evolution}.
The computations required 143 hours to complete 600 time steps (15,200 s of simulated time). \textcolor{black}{Adaptive mesh refinement and coarsening was used to reduce the necessary computation time. However, while mesh refinement preserves the property of mass conservation imposed by the governing equations, the mesh coarsening algorithm can introduce small changes in the global mass. This loss of mass conservation in the mesh coarsening leads to slight fluctuations in the energy evolution (see Figure \ref{fig:pf_energy}).}

\subsubsection{Multiple precipitates}
To study the relative position of equilibrium precipitates, eight spherical precipitates were seeded in a Mg matrix with a domain of $80 \times 80 \times 220$ nm$^3$. These precipitates were also initialized with $c_{\mathrm{p}_0} = 0.125$ and an average Y composition of $c_\mathrm{avg} = 0.028$ in the domain. Finite element meshes with up to about 14M degrees of freedom were used. This simulation was run on 420 physical cores over twenty compute nodes on XSEDE Comet.

\begin{figure}[tb]
        \centering
\begin{minipage}[t]{0.24\textwidth}
        \centering
	\includegraphics[width=0.95\textwidth]{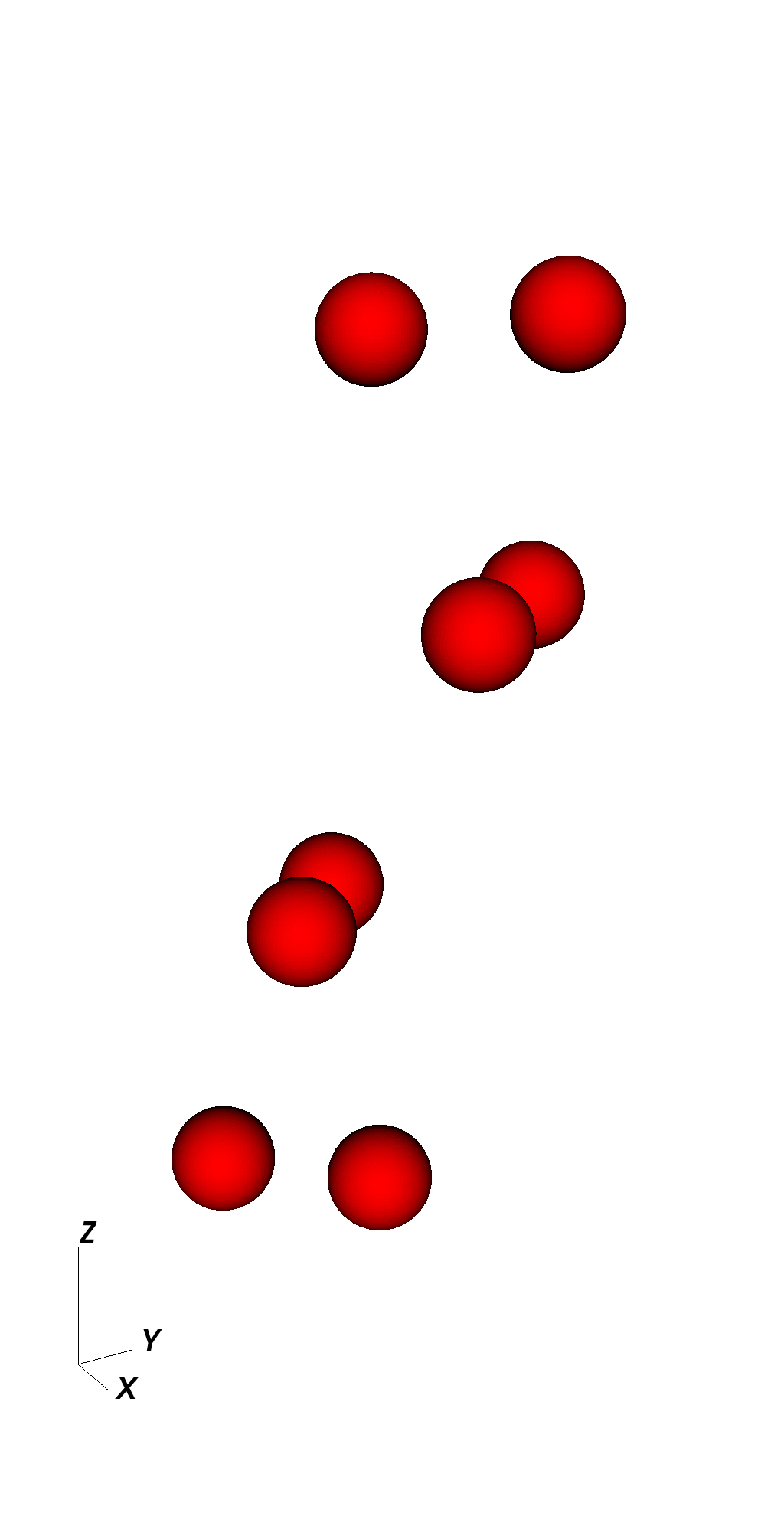}
	\captionof{subfigure}{Initial seeds}
\end{minipage}
\begin{minipage}[t]{0.24\textwidth}
        \centering
	\includegraphics[width=0.95\textwidth]{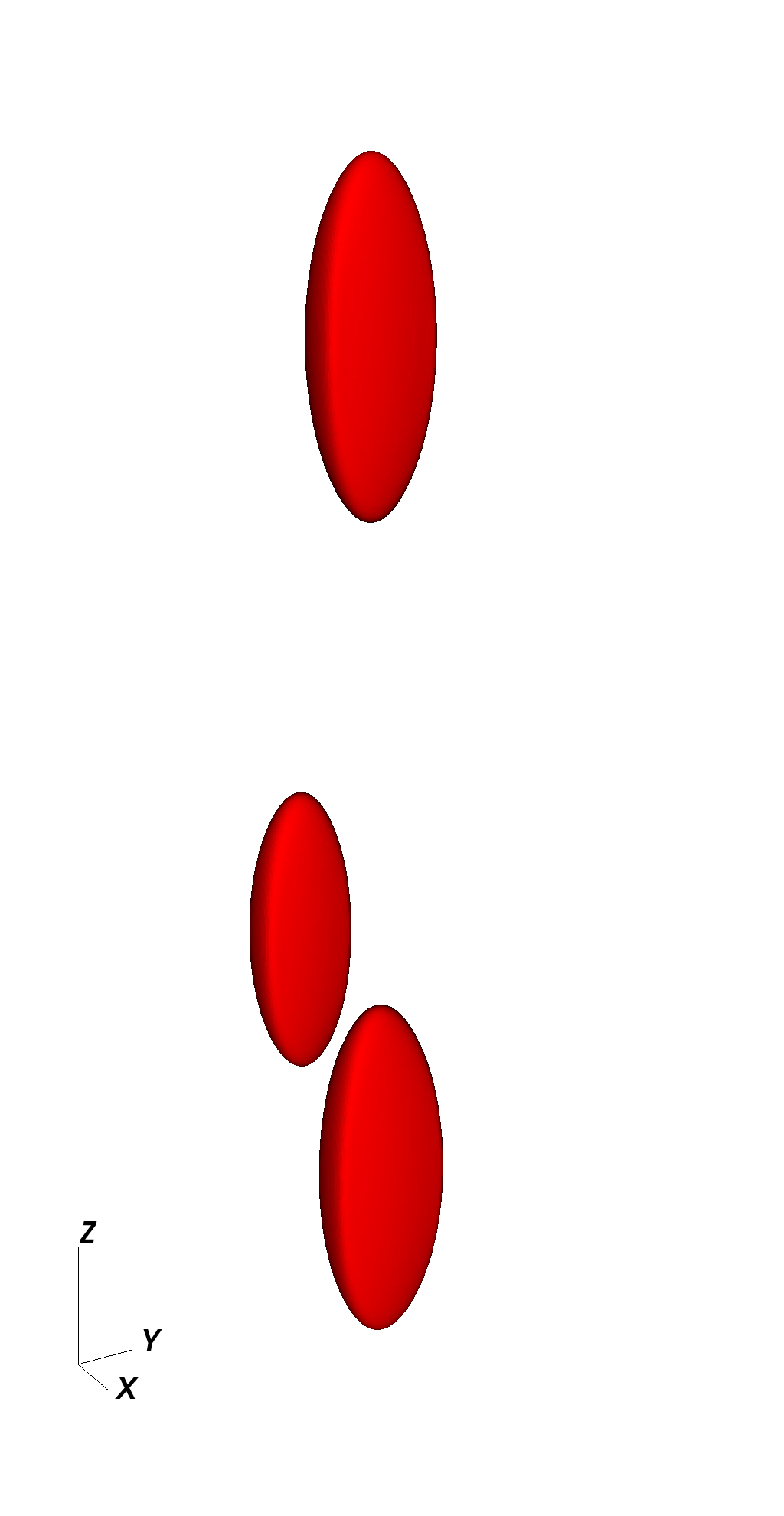}
	\captionof{subfigure}{Time step 300}
\end{minipage}
\begin{minipage}[t]{0.24\textwidth}
        \centering
	\includegraphics[width=0.95\textwidth]{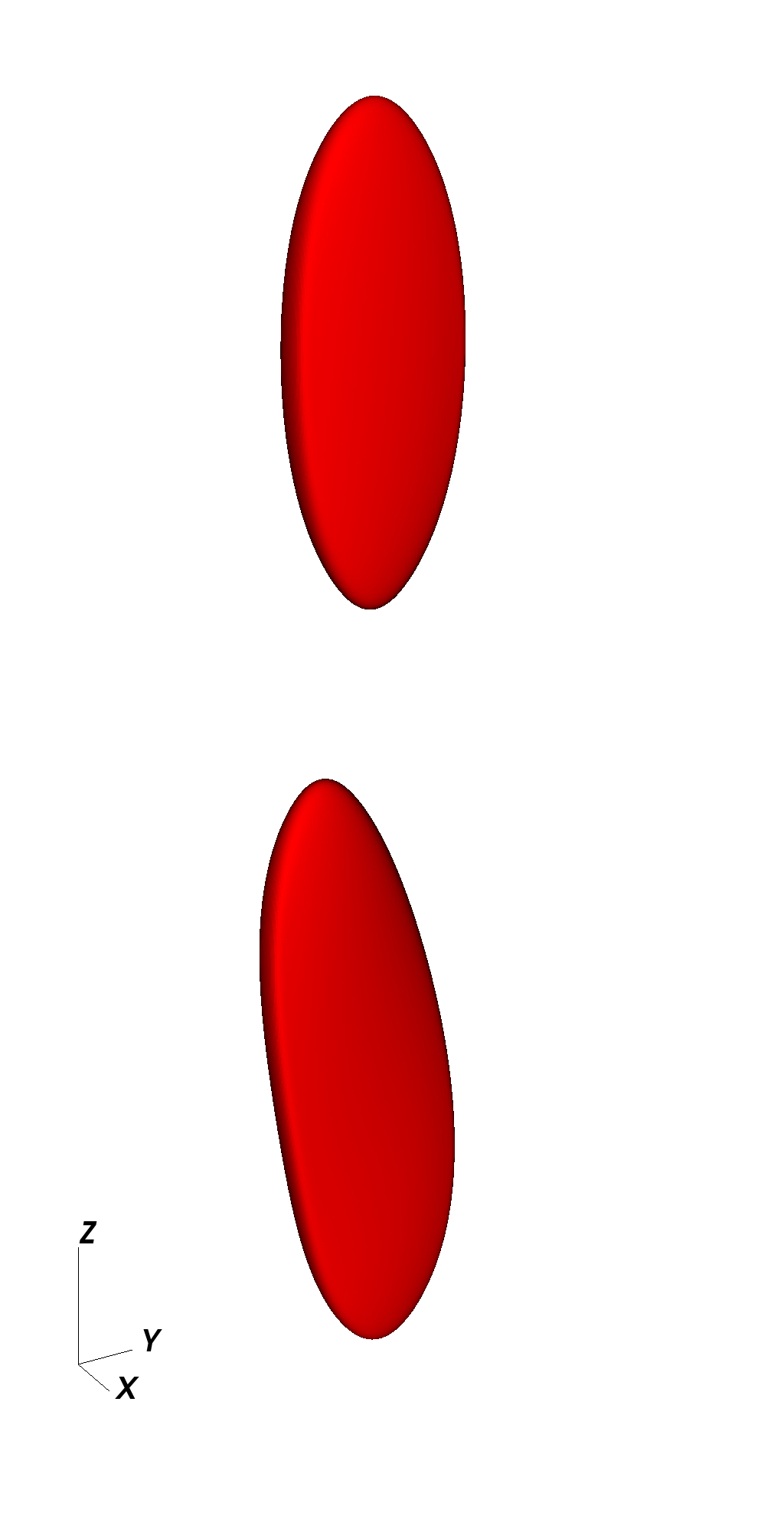}
	\captionof{subfigure}{Time step 600}
\end{minipage}
\begin{minipage}[t]{0.24\textwidth}
        \centering
	\includegraphics[width=0.95\textwidth]{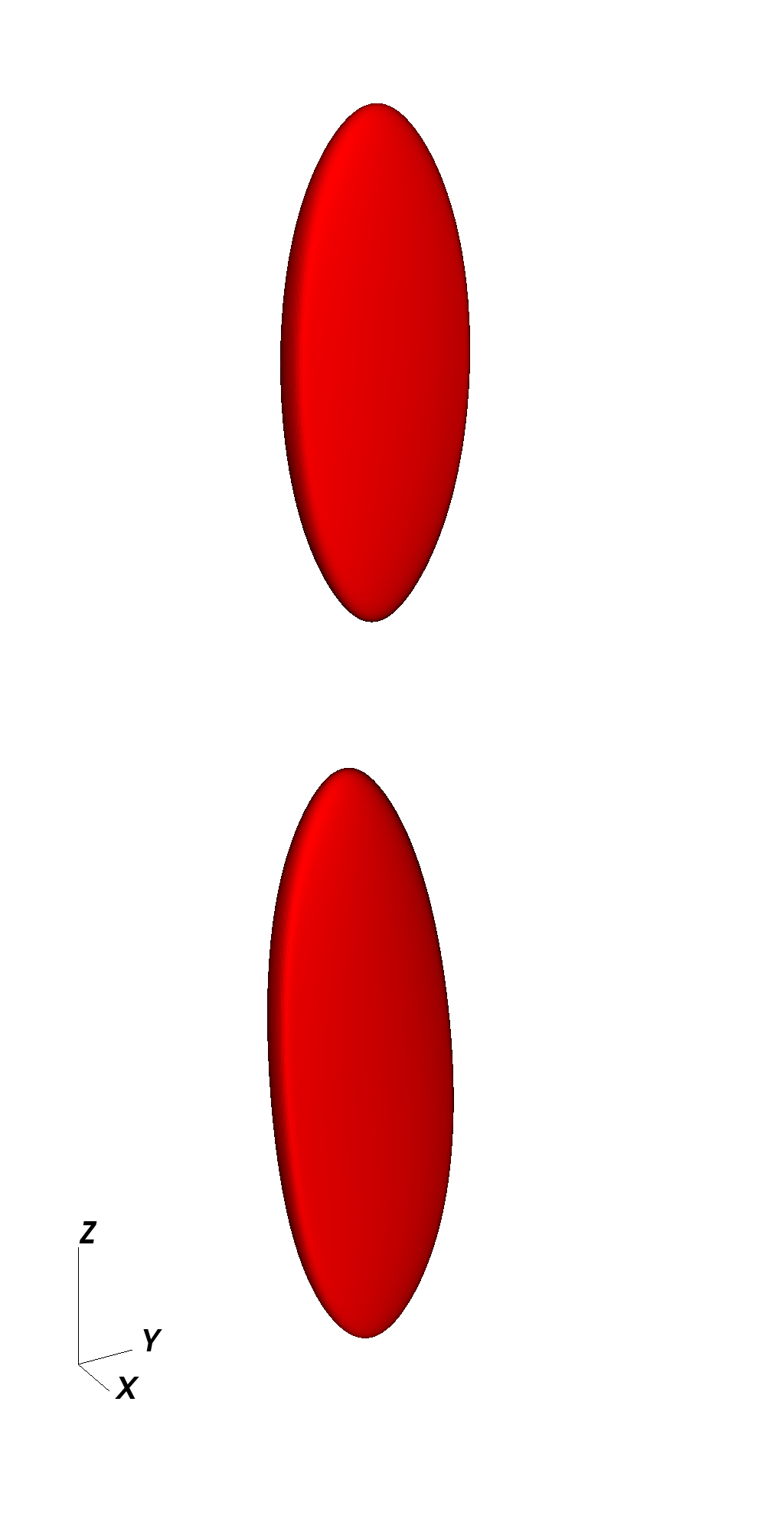}
	\captionof{subfigure}{Time step 890}
\end{minipage}
        \caption{Growth and evolution of the eight precipitate seeds over time as modeled by the phase field method.}
	\label{fig:PFmulti_evolution}
\end{figure}

\begin{figure}[tb]
        \centering
        \includegraphics[width=0.55\textwidth]{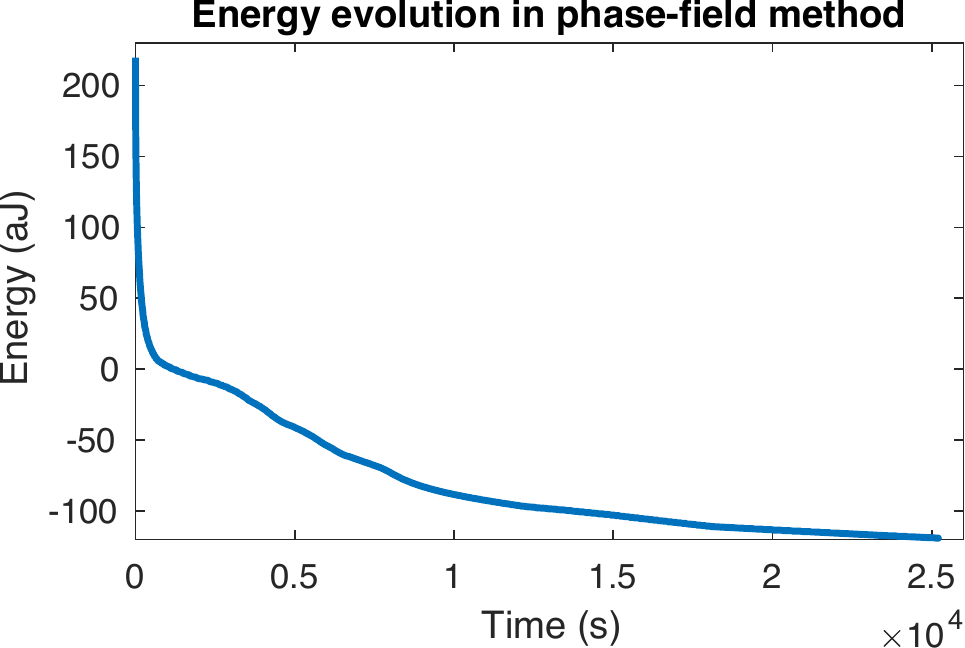}
        	\caption{Evolution of the total energy in the phase field method over simulated time for multiple precipitate seeds.}
	\label{fig:pfmulti_energy}
\end{figure}

The phase field solution was approaching an equilibrium after 890 time steps. 
\textcolor{black}{Depending on location, some of the initial seeds grew smaller and vanished, while others grew larger and merged, leaving two precipitates. It is interesting to note that the two precipitates do not appear to be exactly the same size or have perfect symmetry. However, further small changes may occur if the simulation were to continue. The simulation was stopped before reaching a complete equilibrium due to the effect of adaptive mesh refinement and coarsening on mass conservation.} The evolution of the phase field solution over time appears in Figure \ref{fig:PFmulti_evolution}.
The computations required 224 hours to complete 890 time steps (about 25,000 s of simulated time). Again, the mesh coarsening due to the adaptive mesh refinement and coarsening algorithm does not conserve mass, and periodic jumps are observed in the energy evolution (see Figure \ref{fig:pfmulti_energy}).

\subsection{Discussion}
As observed in the results of the previous section, the computation time required to reach an equilibrium state using phase field methods can be significant. Phase field methods impose first-order dynamics to traverse the free energy landscape toward minima, which are states corresponding to equilibrium precipitate shapes. Such approaches have the potential disadvantages of becoming trapped in local minima, or undergoing slowly evolving dynamics. The dynamics can become stiff due to sharp transitions between states of the system, leading to difficulties with stability and convergence. Additionally, the dynamics inherent in phase field methods evolves in a serial manner. As a counterpoint, this work studies the use of machine learning methods to represent the free energy surface in a relatively low-dimensional space, and find optimal states (minima). As opposed to the serial nature of phase field dynamics, the generation of data from direct numerical simulations to create a free energy surface can be carried out in a massively parallel manner. Knowledge of the landscape can accelerate its traversal and detection of minima using machine learning, as we now describe.

\section{Machine learning based shape prediction}
\label{sec:DNS-ML}

We approach the problem of precipitate shape prediction as one of energy minimization over possible shapes and compositions. The approach combines data generation by direct numerical simulation (DNS) with machine learning (DNS-ML). In the following sections, we first present the methods used to represent precipitate shape and compute its total energy by DNS. Next, we describe the surrogate-based optimization method, including sampling methods, multifidelity modelling using Deep Neural Networks (DNNs), and sensitivity analysis. The phase field results are then compared with the DNS-ML method.

\subsection{Reduced-order model for precipitate shape representation}
\label{sec:ROM}

Precipitate shapes can be represented by a small number of parameters, which become features in the ML models. The phase field models of Section \ref{sec:phase field} include millions of finite element degrees of freedom (DOFs). Of these, $\sim 10^5$ composition DOFs are needed to resolve the precipitate-matrix interface. The approach developed in this section relies on DNS for data generation, which also demands similarly high resolution of the precipitate-matrix mesh as the phase field computations. However, the ML component of our approach offers opportunities for reduced-order models to represent the precipitate. To use a small number of shape parameters while maintaining a large degree of generality in the potential shapes, we use a quadratic B-spline surface with eleven free parameters to model the precipitate shape. These parameters are included in the ML feature vectors.

The quadratic B-spline surface is defined using 20 control points (see Table \ref{tab:controlPoints}) and knot vectors $U = (0,0,1,2,3,3,3)$ and $V = (0,1,2,3,4,5,6,7)$.
\textcolor{black}{After incorporating the desired smoothness and symmetry, the number of free shape parameters defining the location of the control points reduces from sixty ($3\times20$) to eleven.}
Given an orthonormal Euclidean basis set $\{\bsym{e}_1,\bsym{e}_2,\bsym{e}_3\}$, a precipitate with its centroid at $\bsym{X} = \bsym{0}$ possesses symmetry about each of the planes with normal $\pm\bsym{e}_i,\;i = 1,2,3$. It is therefore sufficient to represent only one-eighth of the surface. The first three parameters, $a$, $b$, and $c$, give the bounding dimensions of the surface. The surface is centered on $\bsym{X} = \bsym{0}$, and each of the points $\bsym{X} = a\bsym{e}_1,\; b\bsym{e}_2, c\bsym{e}_3$ lies on the surface. The eight remaining parameters, $t_1,\dots t_8 \in [0,1]$ further define the surface topography.

\begin{table}[tb]
    \centering
    \caption{Control points describing the precipitate surface.}
    \begin{tabular}{c c c}
    $i$ & $j$ & $\bsym{B}_{ij}$\\
    \hline
    1 & 1 & $(-t_1a,b,-t_5c)$\\
    1 & 2 & $(t_1a,b,-t_5c)$\\
    1 & 3 & $(a,t_2b,-t_6c)$\\
    1 & 4 & $(a,-t_2b,-t_6c)$\\
    1 & 5 & $(t_1a,-b,-t_5c)$\\
    2 & 1 & $(-t_1a,b,t_5c)$\\
    2 & 2 & $(t_1a,b,t_5c)$\\
    2 & 3 & $(a,t_2b,t_6c)$\\
    2 & 4 & $(a,-t_2b,t_6c)$\\
    2 & 5 & $(t_1a,-b,t_5c)$\\
    3 & 1 & $(-t_3t_8a,t_7b,c)$\\
    3 & 2 & $(t_3t_8a,t_7b,c)$\\
    3 & 3 & $(t_8a,t_4t_7b,c)$\\
    3 & 4 & $(t_8a,-t_4t_7b,c)$\\
    3 & 5 & $(t_3t_8a,-t_7b,c)$\\
    4 & 1 & $(0,0,c)$\\
    4 & 2 & $(0,0,c)$\\
    4 & 3 & $(0,0,c)$\\
    4 & 4 & $(0,0,c)$\\
    4 & 5 & $(0,0,c)$
    \end{tabular}
    \label{tab:controlPoints}
\end{table}

The parametric function describing the quadratic B-spline surface over one-eighth of the precipitate is given by
\begin{equation}
    \bsym{r}(u,v) = \sum_{i=1}^4\sum_{j=1}^5 \bsym{B}_{ij}N_{i,2}(u)M_{j,2}(v)
    \label{eqn:splinesurfparam}
\end{equation}
where \textcolor{black}{$u\in[1,3]$, $v\in[2,4]$}, and $N_{i,p}$ is the B-spline basis function of order $p$. The basis functions are defined by the Cox-de Boor recursion formula
\begin{align}
    N_{i,p}(u) &= \frac{u - u_i}{u_{i+p}-u_i}N_{i,p-1}(u)+\frac{u_{i+p+1} - u}{u_{i+p+1}-u_{i+1}}N_{i+1,p-1}(u)\\
    N_{i,0}(u) &= 
    \begin{cases}
        1 & \text{if } u_i \leq u < u_{i+1}\\
        0 & \text{otherwise}
    \end{cases}
\end{align}
using the knot vector $U = \{u_1,u_2,\ldots,u_{n+p+1}\}$. $M_{j,p}(v)$ is similarly defined using the knot vector $V = \{v_1,v_2,\ldots,v_{m+p+1}\}$.

\subsection{Energy data from Direct Numerical Simulation}
\label{sec:DNS}
The equilibrium precipitate shapes depend on the strain, interfacial, and chemical energies. The total energy, given by the following integral, is minimized for a given precipitate volume:
\begin{equation}
\Pi = \int_{\Omega_0} \left(\psi + f\right) \,\mathrm{d}V + \int_{\Gamma_0} \gamma(\bsym{n})\,\mathrm{d}S
\label{eqn:DNS_energy}
\end{equation}
where $\Gamma_0$ is the precipitate-matrix interface with unit normal $\bsym{n}$, and $\gamma(\bsym{n})$ is the orientation-dependent interfacial energy discussed in Section \ref{sec:gradientenergy}.

\subsubsection{Strain energy}
\label{sec:MLstrain}

\begin{figure}[tb]
        \centering
        \includegraphics[width=0.45\textwidth]{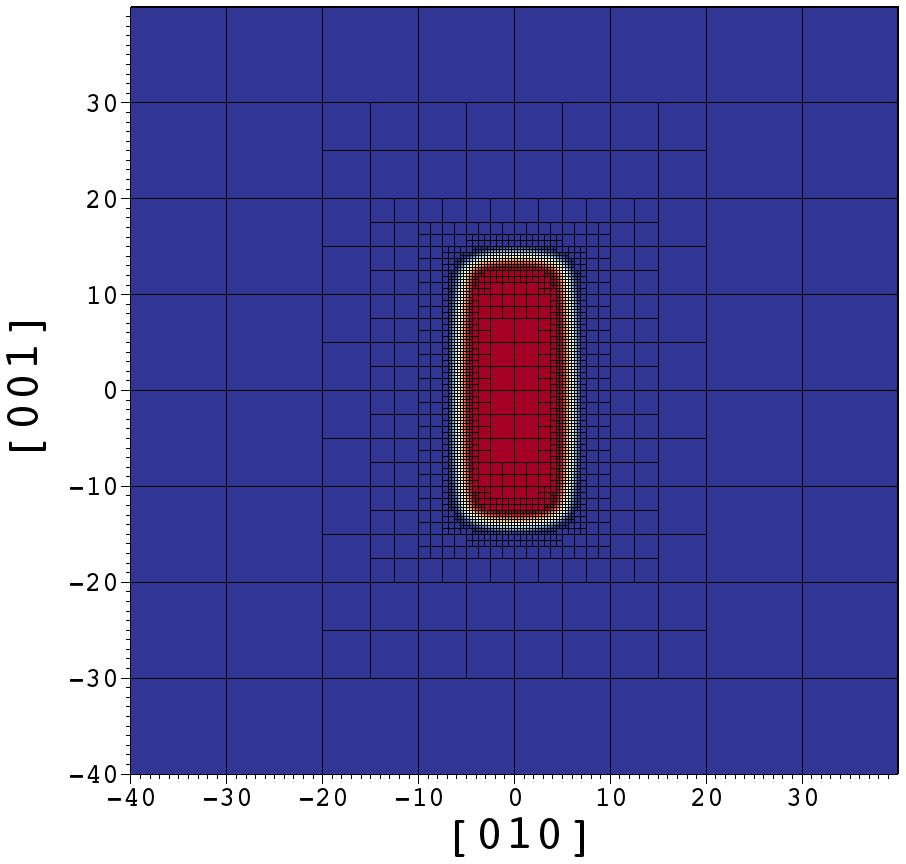}
        	\caption{2D slice of a high-fidelity DNS result showing the finite element mesh.}
	\label{fig:dns_mesh}
\end{figure}

The strain energy of the precipitates was computed by solving the finite strain, continuum elasticity problem by the finite element method, using C++ code\footnote{The code is also available upon request and will soon be released as an application in the {\tt mechanoChemFEM} library described previously.} based on the {\tt deal.II} library. The same domain and a similar mesh refinement were used as in Section \ref{sec:phase field}. The predefined precipitate-matrix interface is represented as a parametric surface, $\br(u,v)$. A signed distance function, $\chi(\bsym{X})$, with positive values inside the precipitate, can be constructed using $\br(u,v)$. A structured finite element mesh was used that had been locally refined at the precipitate-matrix interface using hanging nodes (see, for example, Figure \ref{fig:dns_mesh}). The discontinuity in material parameters at the precipitate-matrix interface, $\chi = 0$, was smoothed linearly over multiple elements, representing a diffuse interface of thickness $\delta$ by defining \textcolor{black}{a second regularized Heaviside function:}
\begin{align}
    \widetilde{H}(\bsym{X}) &=
    \left\{\begin{array}{cl}
    0,& \chi(\bsym{X}) < -\delta/2 \\
    1,& \chi(\bsym{X}) > \delta/2 \\
    \chi(\bsym{X})/\delta + 1/2,& \text{otherwise}
    \end{array}
    \right.\label{eqn:Heaviside}
\end{align}

The displacement field is driven by the eigenstrain, resulting from the lattice parameter mismatch between the precipitate and the matrix. The value of the misfit strain is dependent on the composition of the precipitate \cite{Natarajan2017}. The Mg-Y $\beta'$ precipitate has a composition of Mg$_7$Y, corresponding to $c_\mathrm{Y} = 0.125$ (see Table \ref{tab:eigenstrain}). The eigenstrain is applied within the precipitate using the multiplicative decomposition of the deformation gradient, $\bsym{F} = \bsym{F}^\text{e}\bsym{F}^\lambda$ introduced in Equation (\ref{eqn:FeFlam}) in the context of the phase field method.

\begin{table}[tb]
    \centering
    \caption{Deformation gradient representing the eigenstrain in the Mg-Y $\beta'$ precipitate \cite{Natarajan2017}.}
    \begin{tabular}{c | c }
     & $c_{\mathrm{Y}} = 0.125$\\
    \hline
    $F_{\beta^\prime_{11}}$ & 1.0307\\
    $F_{\beta^\prime_{22}}$ & 1.0196\\
    $F_{\beta^\prime_{33}}$ & 0.9998
    \end{tabular}
    \label{tab:eigenstrain}
\end{table}

The St. Venant-Kirchhoff strain energy density function $\psi(\bsym{E^\text{e}})$, introduced in Equation (\ref{eqn:SVK}) in the phase field setting, is used. The elasticity tensor, with linear spatial variation over the diffuse interface, is given by

\begin{equation}
    \bbm{C}(\bsym{X}) = \bbm{C}^\alpha\cdot\widetilde{H}(\bsym{X}) +  \bbm{C}^{\beta^\prime}\cdot(1-\widetilde{H}(\bsym{X}) ),
 \label{eqn:Calphabeta}
\end{equation}
The strain energy density is numerically integrated over $\Omega_0$ to compute the total strain energy.

A single precipitate is simulated in a Mg matrix with a domain of $80 \times 80 \times 110$ nm$^3$. Dirichlet conditions are imposed; the normal components of the displacement field  are constrained to vanish on the boundaries of the cube. The boundary value problem of nonlinear elasticity, introduced in Equation (\ref{eqn:elastweakform}) is solved.

\subsubsection{Interfacial energy}
The interfacial energy per unit area at a point on the surface of the precipitate is dependent on the orientation of the surface normal. This anisotropic interfacial energy, $\bsym{\gamma}$, and the unit normal, $\bsym{n}$, are used to find the interfacial energy per unit area. The total interfacial energy, $\Pi_{\Gamma_0}$, is found by numerically integrating this expression over the parametric surface of the precipitate:
\begin{equation}
\Pi_\Gamma = 
\int_3^4 \int_2^3 \bsym{n}\cdot\bsym{\gamma}\bsym{n} \left\lVert\frac{\partial\br}{\partial u}\times\frac{\partial\br}{\partial v}\right\rVert\,\mathrm{d}u\mathrm{d}v
\label{eqn:DNS_IE}
\end{equation}
where $\bsym{r}$ is the position vector and the diagonal tensor $\bsym{\gamma}$ was introduced in Table \ref{tab:interfacialEnergy}.

\subsubsection{Bulk chemical free energy}
At equilibrium, we take the composition of the precipitate, $c_\mathrm{p}$, and the composition of the matrix, $c_\mathrm{m}$, to be uniform within their respective domains, with a linearly smoothed interface. The average composition imposed over the domain is  $c_\text{avg} = 0.28$. We take the concentration, $c$, at $\bsym{X}$ to be given by

\begin{equation}
    c = c_\text{p}\widetilde{H}(\bsym{X}) +  c_\text{m}(1-\widetilde{H}(\bsym{X}) )
 \label{eqn:calphabeta}
\end{equation}
Then, given the total volume of the  domain, $V$, the composition of the precipitate, and the average composition, the matrix composition, $c_\text{m}$ is
\begin{align}
c_\text{m} &= \frac{1}{V - \int_\Omega \widetilde{H}(\bsym{X})\,\mathrm{d}V}
\left(c_\text{avg}V - c_\text{p}\int_{\Omega_0} \widetilde{H}(\bsym{X}) \,\mathrm{d}V\right)
\end{align}
We compute the bulk chemical free energy as
\begin{equation}
\int_{\Omega_0} \left(f^\alpha(c^\alpha)\left(1-\widetilde{H}(\bsym{X})\right)
+f^{\beta'}(c^{\beta'})\widetilde{H}(\bsym{X}) \right)\,\mathrm{d}V
\end{equation}
where $f^{\beta'}$ and $f^{\alpha}$ are the chemical free energy energy densities of the $\beta'$ precipitate and the $\alpha$-Mg matrix, respectively.

The functions $f^\alpha(c^\alpha)$ and $f^{\beta'}(c^{\beta'})$ are defined by Equations (\ref{eqn:f_alpha}) and (\ref{eqn:f_beta}), introduced for the phase field formulation in Section \ref{sec:phase field}. The equations for $c^\alpha$ and $c^{\beta'}$ are similar to those in Equations (\ref{eqn:calpha}) and (\ref{eqn:cbeta}):
\begin{align}
c^\alpha &= \frac{A^{\beta'}\left[c - c^{\beta'}_0\widetilde{H}(\bsym{X})\right] + A^\alpha c^\alpha_0\widetilde{H}(\bsym{X})}{A^\alpha \widetilde{H}(\bsym{X}) + A^{\beta'}(1 - \widetilde{H}(\bsym{X}))}\\
c^{\beta'} &= \frac{A^\alpha\left[c - c^\alpha_0(1-\widetilde{H}(\bsym{X}))\right] + A^{\beta'} c^{\beta'}_0(1-\widetilde{H}(\bsym{X}))}{A^\alpha \widetilde{H}(\bsym{X}) + A^{\beta'}(1 - \widetilde{H}(\bsym{X}))}
\end{align}

\subsection{Surrogate based optimization}
\label{sec:surrogateOpt}

The total energy of the system is minimized using surrogate based optimization. Surrogate based techniques develop proxy models of data for carrying out optimization. These techniques are useful for introduction of  a mathematical representation for the evaluation of gradients of the surface being traversed. 
\textcolor{black}{Since we expect the data points to lie on a smooth energy surface, a surrogate based approach is appropriate. Simulated annealing or genetic algorithms would also be valid approaches, in that they do not require a gradient. They would be particularly useful if features with discrete values, such as the number of facets, were considered.}

Here, the surrogate is a multifidelity model representing the free energy representation, derived from corresponding data values. We refer to the work of Vu and co-workers \cite{Vu2017} who describe the following algorithm:

\begin{algo}
Surrogate-based optimization

\fbox{\begin{minipage}{11cm}
{\tt
\begin{enumerate}
    \item Phase 1 (design): Let $k := 0$. Select and evaluate a set $S_0 = \{\boldsymbol{\Xi}_{1_0},\dots\boldsymbol{\Xi}_{p_0}\}$ of starting points,$\boldsymbol{\Xi}_{1_0},\dots\boldsymbol{\Xi}_{p_0}\in\mathbbm{R}^n$. The outputs are $\{\Upsilon_{1_0},\dots\Upsilon_{p_0}\}$, $\Upsilon_{1_0},\dots\Upsilon_{p_0} \in \mathbb{R}$.
    \item While some given stopping criteria are not met:
    \begin{enumerate}
        \item Phase 2 (model): From the data $\{(\boldsymbol{\Xi}_{q_k}, \Upsilon_{q_k}) \, | \, \boldsymbol{\Xi}_{q_k} \in S_k \}$, construct a surrogate model $s_k(\cdot)$ that approximates the black-box function.
        \item Phase 3 (search): Use $s_k(\cdot)$ to search for a new point to evaluate. Evaluate the new chosen point, update the data set $S_k$. Assign $k := k + 1$.
    \end{enumerate}
\end{enumerate}}
\end{minipage}}
\label{algo:surrogateopt}
\end{algo}

A brief overview of our application of this algorithm is as follows:

\textcolor{black}{In the first phase, we identify a range of potential values for the features defining the precipitate shape and composition. We use a Sobol$^\prime$ sequence to choose an initial set of precipitate shapes and compositions from this domain of interest. We compute the total energies for this initial set. In phase 2, we use the current set of DNS energy data to train a multifidelity model based on DNNs. This model gives an approximation of the energy landscape related to precipitate morphology. In the last phase, The multifidelity energy model is minimized, giving an estimated equilibrium precipitate shape and composition. A sensitivity analysis of the model is performed. The domain of interest for the most influential features is tightened around the current estimated minimum. Additional computations are submitted from the updated domain, and the results are added to the DNS energy data set.}
    
In the following sections, we describe further the sampling, multifidelity modeling, minimization, and sensitivity analysis used in the optimization algorithm.

\subsection{Sampling via Sobol$^\prime$ sequences}
\label{sec:sampling}

Surrogate-based optimization involves sampling data points from the domain of the chosen variables. Vu and co-workers outline two important requirements for sampling \cite{Vu2017}:
\begin{enumerate}
    \item Space-fill: The design points should be uniformly spread over the design space.
    \item Noncollapse: Two design points should not share any coordinate value if the important dimensions are unknown \textit{a priori}.
\end{enumerate}

One sampling method that is both space-filling and noncollapsing is based on Sobol$^\prime$ sequences \cite{Sobol1967,Bratley1988,Bessa2017}. They consist of $n$-dimensional points in a unit domain where the first $m$ points are well-distributed for any $m$ and no point projections are coincident. Using a Sobol$^\prime$ sequence allows additional points to be continually added to an initial set of well-distributed points while maintaining the space-filling property of the updated set. Such a sequence was first suggested by Sobol$^\prime$ \cite{Sobol1967} in the context of numerical integrals, but it has since been adopted in optimization schemes. In this work, we generate Sobol$^\prime$ sequences in the Euclidean space, $\mathbbm{R}^n$, of the parameters that determine the energy of the precipitate-matrix system. These parameters, at a minimum, include the precipitate geometry and the average composition. The twelve features considered in our study are listed in Table \ref{tab:range}, along with the initial range of values considered. The set could be extended to precipitate-matrix elasticities and interfacial energies. We use the Sobol$^\prime$ sequence to define the initial set of DNS computations, as well as all subsequent DNS computations performed during each iteration of the optimization algorithm.

\begin{table}[tb]
    \centering
    \caption{Range of possible values considered for the twelve features used in the DNS-ML method.}
    \begin{tabular}{c | c c}
    Feature & min & max\\
    \hline
    $a$ (nm) & 1 & 39\\
    $b$ (nm) & 1 & 39\\
    $c$ (nm) & 1 & 54\\
    $t_1$ & 0 & 1\\
    $t_2$ & 0 & 1\\
    $t_3$ & 0 & 1\\
    $t_4$ & 0 & 1\\
    $t_5$ & 0 & 1\\
    $t_6$ & 0 & 1\\
    $t_7$ & 0 & 1\\
    $t_8$ & 0 & 1\\
    $c_\mathrm{p}$ & 0.12 & 0.13
    \end{tabular}
    \label{tab:range}
\end{table}

\subsection{Multifidelity model using a Knowledge-Based Neural Network (KBNN)}
\label{sec:MF}

Machine learning methods such as DNNs can be used to create a surrogate model based on the computed data. The amount of data available for training, however, is constrained by the computation time required by the DNS. The most expensive component of the total energy to compute is the strain energy, which is calculated using the finite element method. It is possible to reduce the total computation time by using a multifidelity approach. Multifidelity models can achieve high accuracy with reduced computation time by combining highly accurate, expensive data with less accurate, abundant data. The less accurate or low-fidelity data provide overall trends in the model, while the highly accurate or high-fidelity data act to correct the model. Relatively few high-fidelity data points are then needed in comparison to the amount of low-fidelity data. The finite element model for the strain energy lends itself well to a multifidelity approach by using coarse and fine meshes for the low- and high-fidelity data, respectively.

\begin{figure}[tb]
        \centering
        \includegraphics[width=0.7\textwidth]{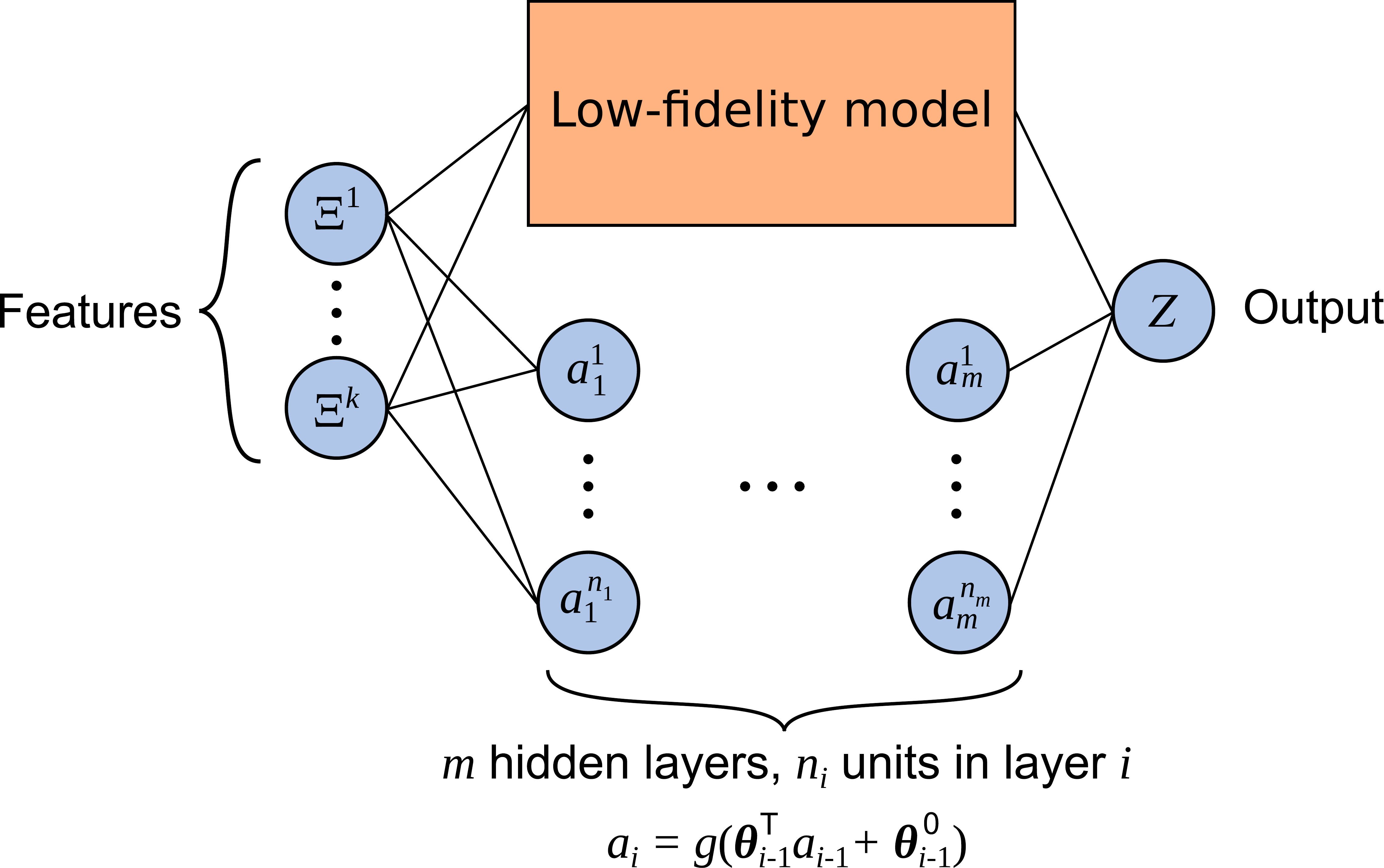}
        	\caption{A Knowledge-Based Neural Network (KBNN) combines a low-fidelity model in parallel with fully-connected neural network layers. \textcolor{black}{In this work, the low-fidelity model used was a standard DNN trained on the low-fidelity data.}}
	\label{fig:KBNN}
\end{figure}

We used a Knowledge-Based Neural Network (KBNN) as the surrogate model to utilize multifidelity data \cite{Leary2003}. A KBNN incorporates some basic knowledge of the system by inserting an analytical low-fidelity model in parallel with the standard, fully-connected layers of a Deep Neural Network (DNN). The KBNN is then trained on the high-fidelity data. The low-fidelity model provides an estimated solution, which the neural network layers improve based on the high-fidelity data. In this work, we use a standard DNN trained to the coarse mesh data points as the low-fidelity model in the KBNN. By using a KBNN to represent the total energy of the precipitate, we are able use coarse mesh computations for most of our data, which significantly reduces the necessary amount of computation time.

The low-fidelity model, $\Pi_L$, is incorporated into the KBNN or high-fidelity model, $\Pi_H$, in the following manner:
\begin{align}
    \Pi_H(\bsym{\Xi}) &= \rho \Pi_L(\bsym{A}\bsym{\Xi} + \bsym{b}) + z(\bsym{\Xi})
\end{align}
where $z(\bsym{\Xi})$ provides corrections to the low-fidelity model, $ \Pi_L$. The coefficient $\rho$ weights the contribution of the low-fidelity model. The diagonal matrix $\bsym{A}$ and the vector $\bsym{b}$ correct $\Pi_L$ by allowing scaling and shifting transformations of the low-fidelity surface within the feature space. The function $z(\bsym{\Xi})$ represents the neural network layers of the KBNN and provides an additive correction that is a function of the feature values. The values of $\rho$, $\bsym{A}$, and $\bsym{b}$, as well as the weights and biases of the neural network layers of $z(\bsym{\Xi})$ are optimized during the KBNN training. This construct, however, allows the possibility for $\rho$ to go to zero during training, thus losing the benefit of the low-fidelity data. To preserve the information given by $\Pi_L$ and assuming necessary shifting and scaling of the inputs to be small, we used the following loss function in training the KBNN:
\begin{align}
    J(\bsym{\Upsilon},\bsym{Z},\rho;\bsym{A};\bsym{b}) = \mathrm{MSE}(\bsym{\Upsilon},\bsym{Z}) + c_1(\rho-1)^2 + c_2||\bsym{A}-\bsym{I}||^2 + c_3||\bsym{b}||^2
\end{align}
where MSE is the mean squared error based on the energy dataset $\bsym{\Upsilon}$ and the corresponding predicted energy $\bsym{Z}$, and $c_1$, $c_2$, and $c_3$ are coefficients that weight the penalty terms.

We used the open source software library TensorFlow \textcolor{black}{\cite{Martin2015}} to create the low-fidelity DNN and the KBNN. TensorFlow's AdagradOptimizer was used as the optimization method in training. The SoftPlus activation function, $g(x) = \ln(1 + e^x)$, was used in the DNN and KBNN. The low- and high-fidelity data points were sorted by energy, and the lowest energy data points were used for training. For iteration $k$, $m_\mathrm{H} = 50(k+1)$ high-fidelity and $m_\mathrm{L} = 2000(k+1)$ low-fidelity data points were used. Mini-batches of 100 data points were used in training. The data were scaled to improve training, with the inputs scaled so that $\Xi^j\in [-1,1]$ and the outputs $Z \in [-10,10]$.

The neural networks are defined by the number of hidden layers and the number of units or neurons in each layer. The optimization function AdagradOptimizer requires a learning rate. The values of these hyperparameters (learning rate, number of hidden layers, and number of units per layer) were determined through a random search \cite{bergstrabengio2012}. The resulting set of hyperparameters was used in the low-fidelity DNN. To prevent overfitting of the KBNN to the relatively low number of high-fidelity data, only two additional neural network layers of 20 neurons each were used in the KBNN.

The hyperparameter search was performed as follows: Given a dataset $S_k$, multiple sets of hyperparameters were created for comparison. These values were randomly selected over the following intervals:
\begin{itemize}
\item learning rate, log-uniformly between $1\times 10^{-4}$ and $0.1$
\item number of hidden layers, log-uniformly between $2$ and $10$
\item units per layer, uniformly between $40$ and $500$
\end{itemize}
The data $\{(\bsym{\Xi}_1,\Upsilon_1),\dots  (\bsym{\Xi}_{p_k},\Upsilon_{p_k})\}$ in dataset $S_k\times \bsym{\Upsilon}_k$ were randomly split into 75\% training data (dataset $S_k^\text{tr}\times \bsym{\Upsilon}^\text{tr}_k$) and 25\% validation data (dataset $S_k^\text{vl}\times \bsym{\Upsilon}^\text{vl}_k$). (Note that $\bsym{\Upsilon}_k \in \bbm{R}^{p_k}$, $\bsym{\Upsilon}^\text{tr}_k \in \bbm{R}^{p^\text{tr}_k}$ and $\bsym{\Upsilon}^\text{vl}_k \in \bbm{R}^{p^\text{vl}_k}$, where $p^\text{tr}_k + p^\text{vl}_k = p_k$.) A neural network defined by each set of hyperparameters was trained with $S_k^\text{tr}$ for $15,000$ iterations of the optimizer. The L$_2$-norm of the error between the output (predicted) and actual (data) values, $Z_l$ and $\Upsilon_l$ ($l = 1,\dots p_k$) for the validation data was computed. The hyperparameter set that produced the lowest error was then used to define the neural network for the current set of data. \textcolor{black}{For the first iteration, 35 random hyperparameter sets were compared, while subsequent iterations compared ten hyperparameter sets.}

\subsection{Minimization of the ML energy surface, and sensitivity analysis}
\label{sec:min}
Once a surrogate model has been constructed, gradient based optimization methods can be used to find the minimum. We again use TensorFlow's AdagradOptimizer to find the minimum of the trained KBNN, with the lowest energy low-fidelity DNS data point as the initial guess. Given a set $\hat{S}_k$ of training data, we define $\mathcal{B}_k = [a_1^\mathrm{min},a_1^\mathrm{max}]\times\ldots\times[a_n^\mathrm{min},a_n^\mathrm{max}]$, where $a_j^\mathrm{min}$ and $a_j^\mathrm{max}$ are the minimum and maximum values of feature $j$ in the set $\hat{S}_k$. Since the KBNN is only expected to be accurate near the data on which it was trained, the minimum was constrained to lie within $\mathcal{B}_k$. The energy surface is explored and refined by submitting additional DNS computations in an updated search space surrounding this predicted minimum. While this minimum of the KBNN is used to guide the submission of additional DNS, its accuracy is not consistent from iteration to iteration. A more stable prediction of the equilibrium precipitate shape at each iteration is the lowest energy high-fidelity data point. It is this more stable prediction that is reported and shown in the results of Section \ref{sec:results}.

We used a global, variance-based sensitivity analysis to determine the relative effect of each of the features on the total energy  \cite{Sobol2001,Saltelli2010}. The sensitivity indices are used, together with the current estimated minimum, to update the set of feature values, $\mathcal{D}_k$, from which simulations will be run in the $k$th iteration of the DNS-ML algorithm. The global sensitivity indices are approximated using a quasi-Monte Carlo algorithm, using tens of thousands of data points. While the computational cost could be prohibitively high if relying only on DNS solutions, the trained multifidelity model can be used to quickly evaluate the necessary data. We compute the total sensitivity index $s^\mathrm{tot}_i$ for each feature, based on the variances $V$ and $V^\mathrm{tot}_i$:
\begin{align}
    s^\mathrm{tot}_i &= \frac{V^\mathrm{tot}_i}{V}\\
    V &\approx \left(\frac{1}{N}\sum\limits_{j=1}^N Z^2(\bsym{\Xi}_j)\right) - \left(\frac{1}{N}\sum\limits_{j=1}^N Z(\bsym{\Xi}_j)\right)^2\\
    V^\mathrm{tot}_i &\approx \frac{1}{2N}\sum\limits_{j=1}^N\left(Z(\bsym{\Xi}_j) - Z({\bsym{\Xi}'}_j^{(i)}) \right)^2
\end{align}
where $\bsym{\Xi}_j$ and ${\bsym{\Xi}'}_j^{(i)}$ are quasi-random points taken from the Sobol$^\prime$ sequence, such that all except the $i$-th components of $\bsym{\Xi}_j$ and ${\bsym{\Xi}'}_j^{(i)}$ are equal. 

A low sensitivity index for feature $i$ implies that the component $\Xi^\mathrm{min}_{k_i}$ of the current minimizer could be far from the true value with little effect on the predicted minimum energy. Thus, we do not reduce the dimension of $\mathcal{D}_{k+1}$ for this feature in the current iteration. Conversely, if feature $i$ has a high sensitivity index, it has a large effect on the energy. Then, the value of component $\Xi^\mathrm{min}_{k_i}$ corresponding to a minimum energy is likely to be close to the true minimizing value of the feature. Therefore, the domain for this feature is tightened around the current prediction. This more localized domain defined by $\mathcal{D}_{k+1}$ is used with the Sobol$^\prime$ sequence to create additional DNS data points for the next iteration of the optimization algorithm. If $m_\mathrm{H}$ and $m_\mathrm{L}$, with $m_\mathrm{H} < m_\mathrm{L}$, are the respective number of new high- and low-fidelity simulations run in each iteration, then the first $m_\mathrm{H}$ sets of feature values in the Sobol$^\prime$ sequence for $\mathcal{D}_{k+1}$ are submitted as both high- and low-fidelity DNS. The following points, up to $m_\mathrm{L}$, are submitted as only low-fidelity DNS.

\textcolor{black}{This scheme for tightening the search space around a predicted minimum is best suited for the case of a clear global minimum. Difficulties may arise when there are multiple energy wells at or near the same depth as the global minimum. The DNS-ML method will always choose a single point to focus around at each iteration. However, the sorting scheme used to select training points will choose points clustered around all low energy wells, at least initially. This could result in the algorithm jumping between energy wells as the iterations progress. As such, it is best to remove regions of the search space that, due to symmetry, would introduce additional minima. For example, while Mg-Y precipitates have multiple rotational variants, only one orientation is considered in the simulations in this work, thus avoiding the multiple energy wells associated with precipitate orientation.}

\section{A DNS-ML algorithm that exploits a heterogeneous computing architecture}
\label{sec:algo}

An algorithm is presented in this section that explores the tight coupling of direct numerical simulations, data generation and machine learning to predict equilibrium precipitate shapes. The workflow is managed by a Python script and executed on the ConFlux High Performance Computing (HPC) cluster at the University of Michigan. The ConFlux cluster includes 58 IBM Power8 CPU ``Firestone'' compute nodes with 20 physical cores (up to 40 virtual cores) each and seventeen additional Power8 CPU ``Minsky'' nodes that each host four NVIDIA Pascal GPUs. All compute nodes and storage are connected using 100 Gb/s InfiniBand fabric. The workflow, machine learning, and optimization routines were implemented in Python and executed on the Minsky nodes, allowing the TensorFlow library to utilize the GPUs during machine learning and optimization. As already outlined, the DNS energy data were computed using finite element code built on {\tt deal.II}, running on the Firestone nodes using 5 virtual cores for each computation. High-speed interconnects enabled rapid transfer of the DNS data from the compute nodes to the GPUs performing the machine learning.

\begin{figure}
    \centering
    \includegraphics[width=0.7\textwidth]{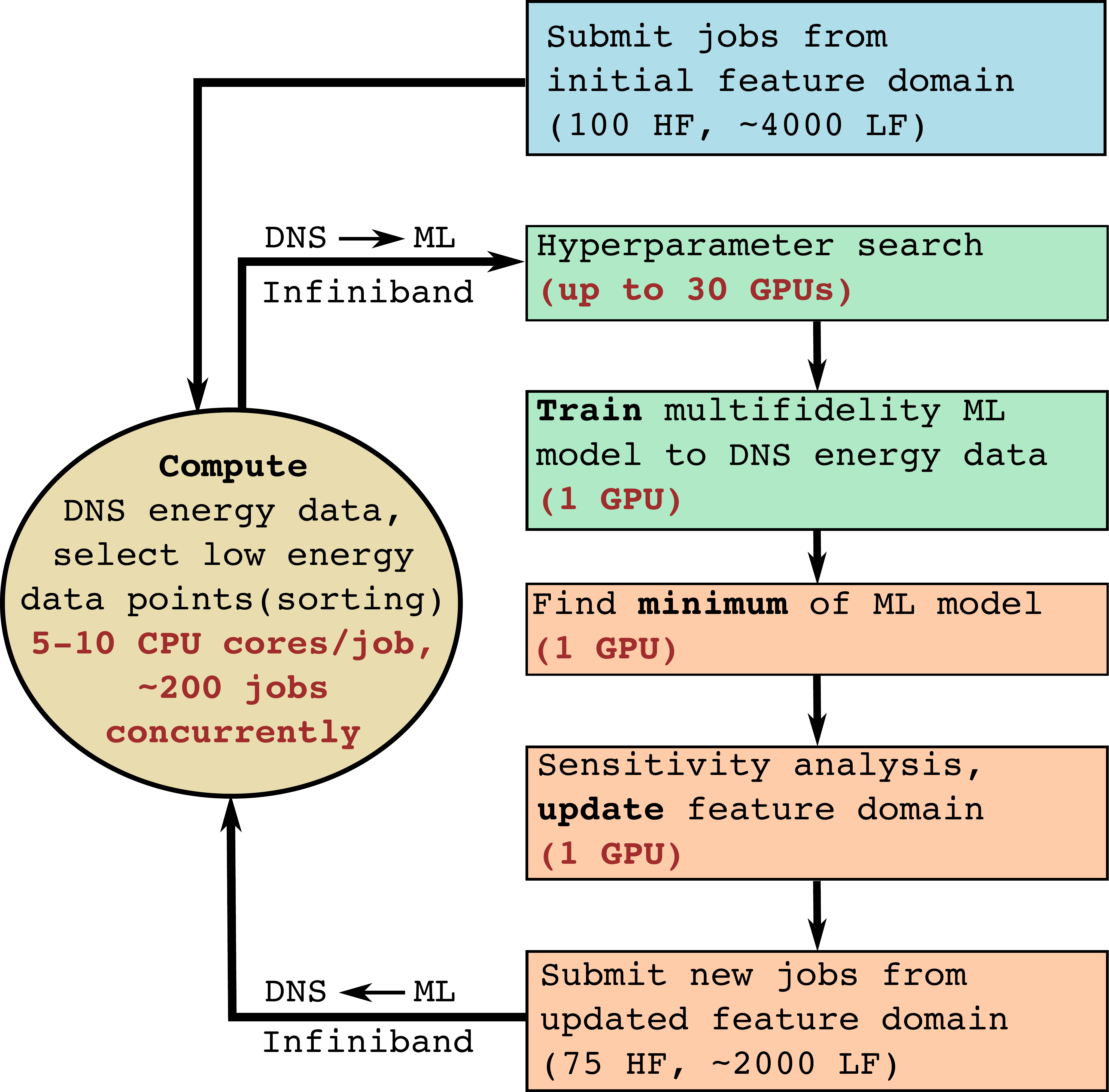}
    \caption{\textcolor{black}{Workflow of the DNS-ML alogrithm.}}
    \label{fig:workflow}
\end{figure}

The following is an expansion of Algorithm \ref{algo:surrogateopt} and a summary of Sections \ref{sec:surrogateOpt}-\ref{sec:min}. The corresponding workflow is shown in Figure \ref{fig:workflow}.

\begin{algo}
DNS-ML algorithm

\fbox{\begin{minipage}{11cm}
{\tt
\begin{enumerate}
    \item Let $k := 0$. Define a set  $\mathcal{D}_0 = [a_{1_0},b_{1_0}]\times\ldots\times[a_{n_0},b_{n_0}]\in\mathbbm{R}^n$ to be the domain of interest. Select and evaluate a set $S_0 = \{\boldsymbol{\Xi}_{1_0},\dots\boldsymbol{\Xi}_{p_0}\}$ of starting points,$\boldsymbol{\Xi}_{1_0},\dots\boldsymbol{\Xi}_{p_0}\in\mathcal{D}_0$ from the Sobol$^\prime$ sequence. The outputs are $\{\Upsilon_{1_0},\dots\Upsilon_{p_0}\}$.
    \item While some given stopping criteria are not met:
    \begin{enumerate}
        \item Sort the points in $S_k$ by $\Upsilon_{q_k}$. Find a subset $\hat{S}_k \subset S_k$ of low energy DNS data. Let $\mathcal{B}_k$ be the smallest Cartesian product such that $\hat{S}_k\subseteq\mathcal{B}_k$.
        \item Train a machine learning model $s_k(\bsym{\xi})$ to the data $\{(\boldsymbol{\Xi}_{q_k}, \Upsilon_{q_k}) \, | \, \boldsymbol{\Xi}_{q_k} \in \hat{S}_k \}$.
        \item Find the point $\bsym{\xi}_\mathrm{min}$ that minimizes $s_k(\bsym{\xi})$ subject to $\bsym{\xi}\in\mathcal{B}_k$.
        \item Compute the global sensitivity indices $s^\mathrm{tot}_i$, $i=1,\ldots,n$ for each feature using $s_k(\bsym{\xi})$.
        \item Tighten the bounds of $\mathcal{D}_k$ based on $\bsym{\xi}_\mathrm{min}$ and $s^\mathrm{tot}_i$ to define $\mathcal{D}_k+1$.
        \item Select an additional set of points $\{\boldsymbol{\Xi}_{1_{k+1}},\dots\boldsymbol{\Xi}_{p_{k+1}}\}$, $\boldsymbol{\Xi}_{1_{k+1}},\dots\boldsymbol{\Xi}_{p_{k+1}}\in\mathcal{D}_{k+1}$ from the Sobol$^\prime$ sequence and compute the outputs. Update $S_k$. Assign $k := k + 1$.
    \end{enumerate}
\end{enumerate}}
\end{minipage}}
\label{algo:surrogateopt2}
\end{algo}

\section{Results}
\label{sec:results}

We present the results of the DNS-ML method for predicting the equilibrium shape of precipitates in Mg-Y and compare them with the phase field results. \textcolor{black}{The same level of mesh refinement was used in all high-fidelity DNS and phase field simulations.} \textcolor{black}{The interface thickness $\delta$ of the high-fidelity DNS was chosen to span roughly five elements; this thickness is uniform for the entire interface. The interface thickness in the phase field simulations is not uniform, since it is related to the anisotropic interfacial energy, as in Eqs. (\ref{eqn:kappa}) and (\ref{eqn:lambda}). The minimum interface thickness of the phase field simulations, corresponding to the (010) plane, is set to be equal to value of $\delta$ used in the high-fidelity DNS.}

\subsection{Single precipitate}

\begin{figure}[tb]
        \centering
        \includegraphics[width=0.5\textwidth]{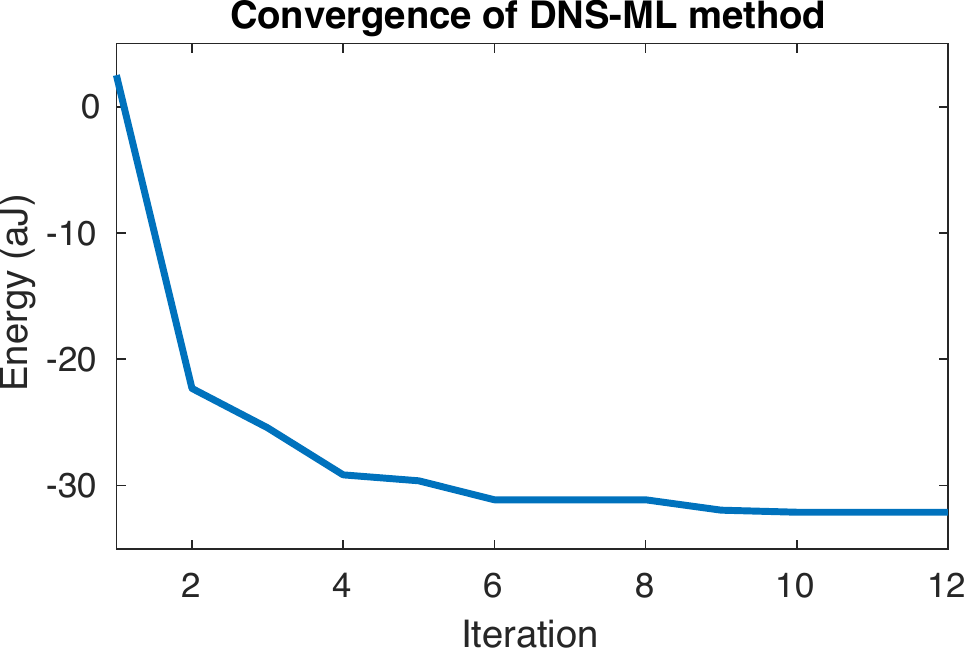}
        	\caption{Convergence of the DNS-ML method for the single precipitate problem, based on the lowest energy high-fidelity DNS computation at each iteration.}
	\label{fig:convergence_cv}
\end{figure}

\begin{figure}[tb]
        \centering
\begin{minipage}[t]{0.24\textwidth}
        \centering
	\includegraphics[width=0.95\textwidth]{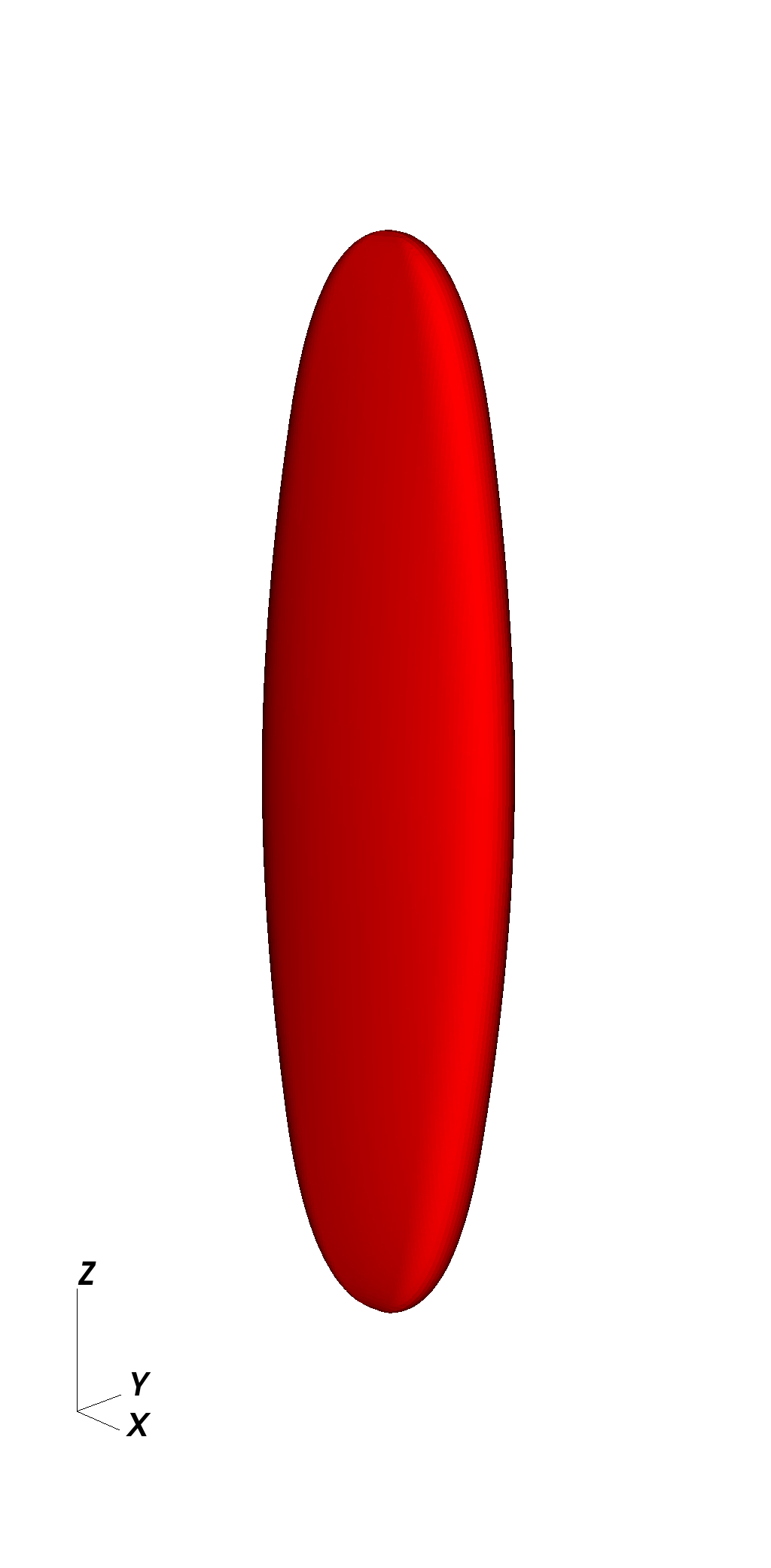}
	\captionof{subfigure}{Iteration 1}
\end{minipage}
\begin{minipage}[t]{0.24\textwidth}
        \centering
	\includegraphics[width=0.95\textwidth]{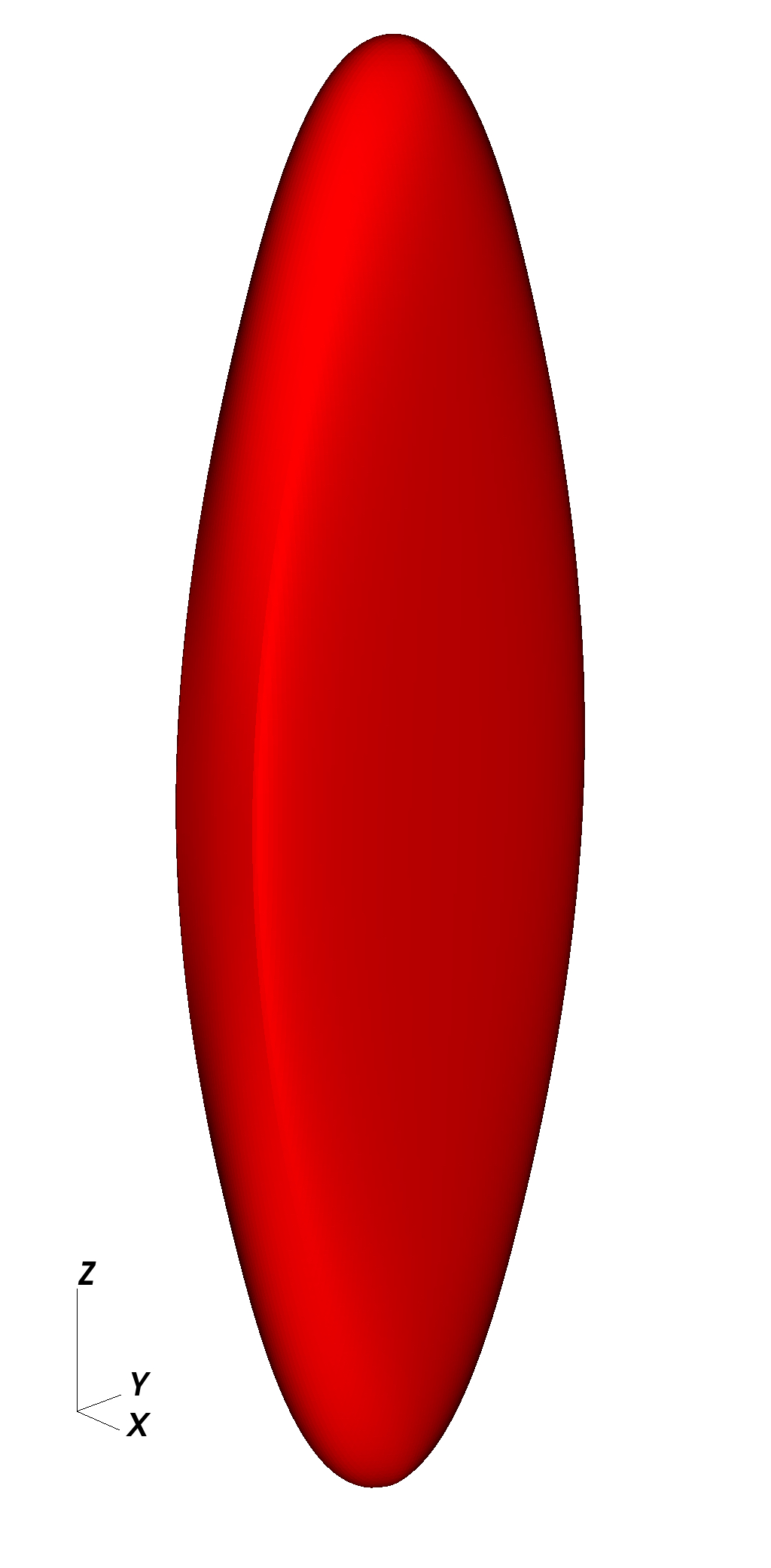}
	\captionof{subfigure}{Iteration 4}
\end{minipage}
\begin{minipage}[t]{0.24\textwidth}
        \centering
	\includegraphics[width=0.95\textwidth]{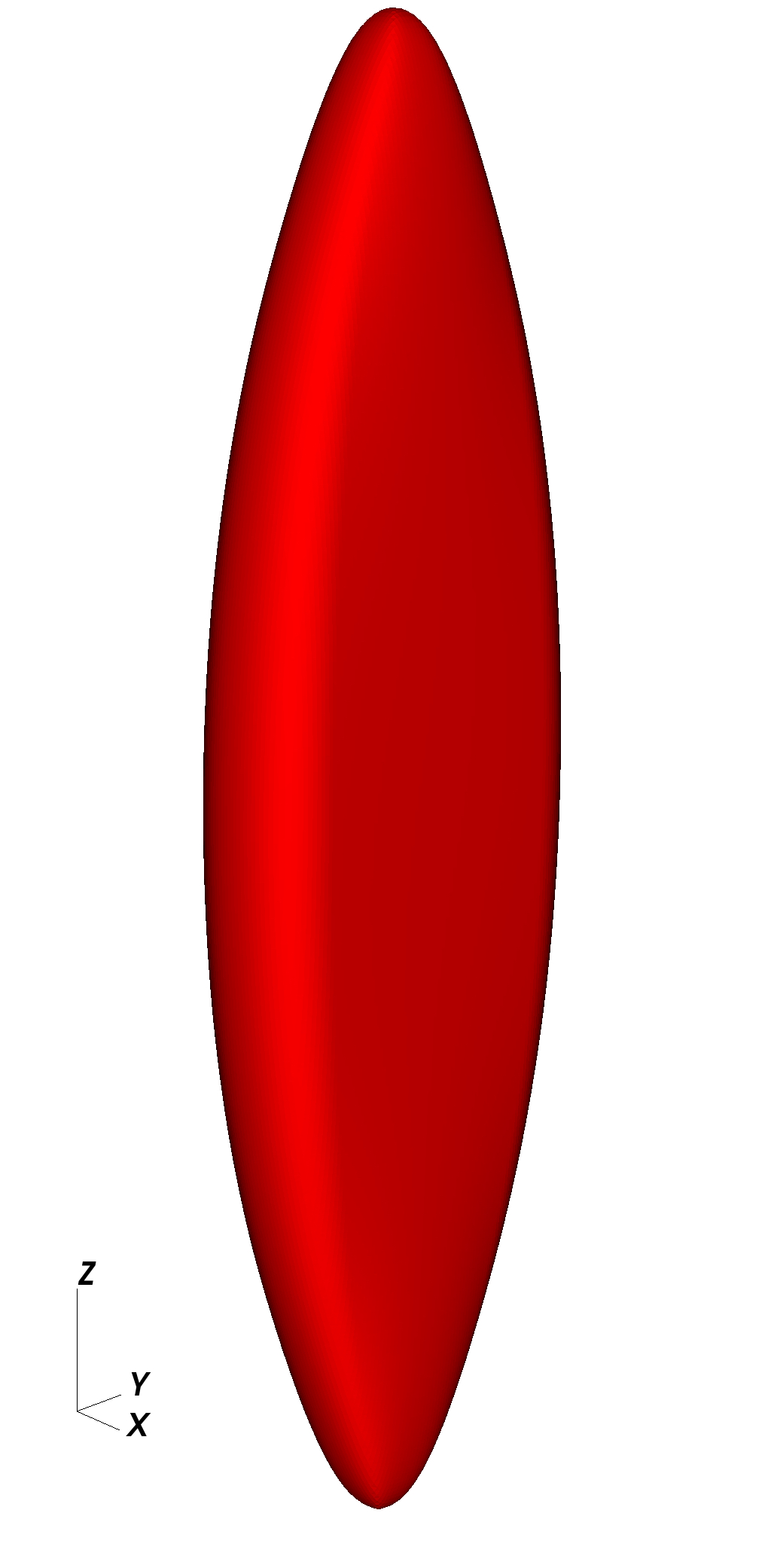}
	\captionof{subfigure}{Iteration 8}
\end{minipage}
\begin{minipage}[t]{0.24\textwidth}
        \centering
	\includegraphics[width=0.95\textwidth]{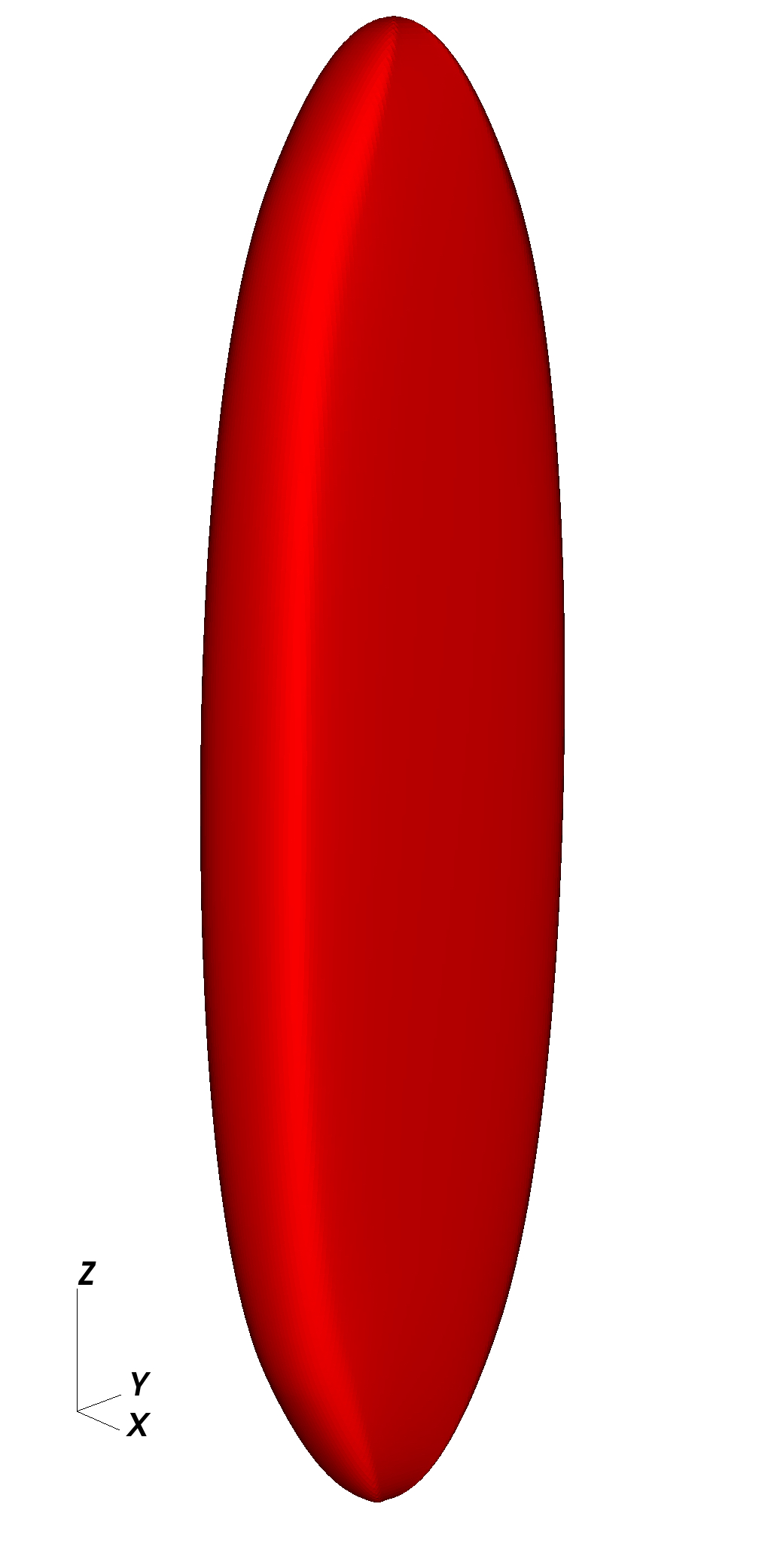}
	\captionof{subfigure}{Iteration 12}
\end{minipage}
        \caption{The predicted precipitate shape at various iterations of the DNS-ML method.}
	\label{fig:prec_DNN}
\end{figure}

\begin{figure}
        \centering
\begin{minipage}[t]{0.4\textwidth}
        \centering
	\includegraphics[width=0.9\textwidth]{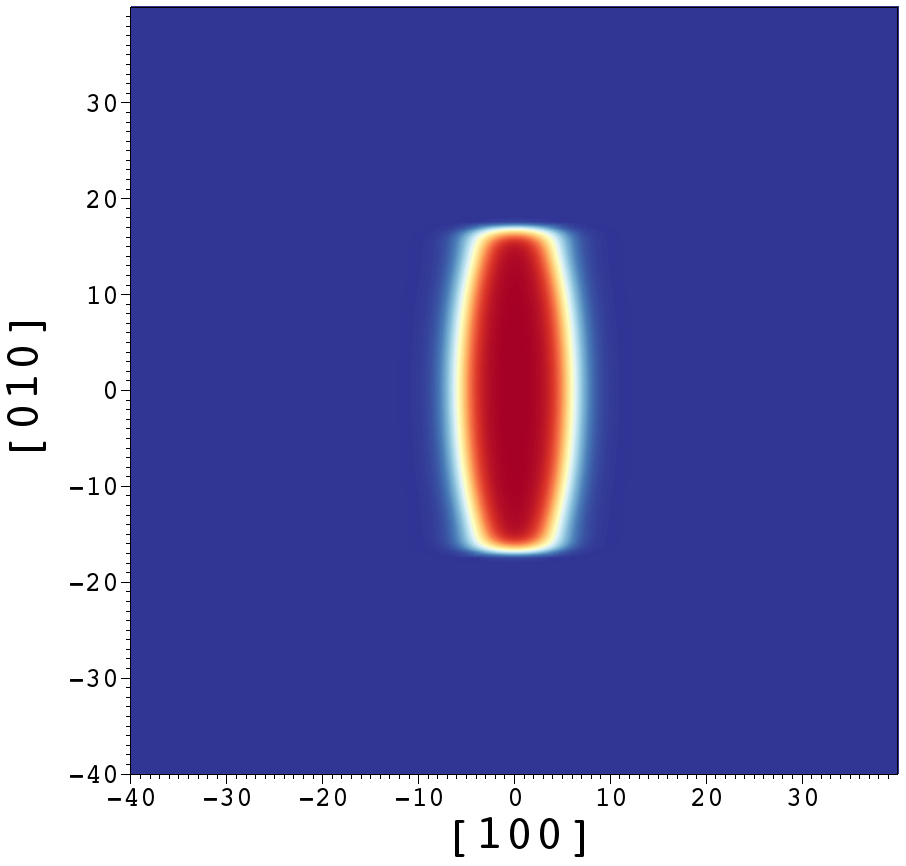}
	\captionof{subfigure}{Phase field, (001) plane}
\end{minipage}
\begin{minipage}[t]{0.4\textwidth}
        \centering
	\includegraphics[width=0.9\textwidth]{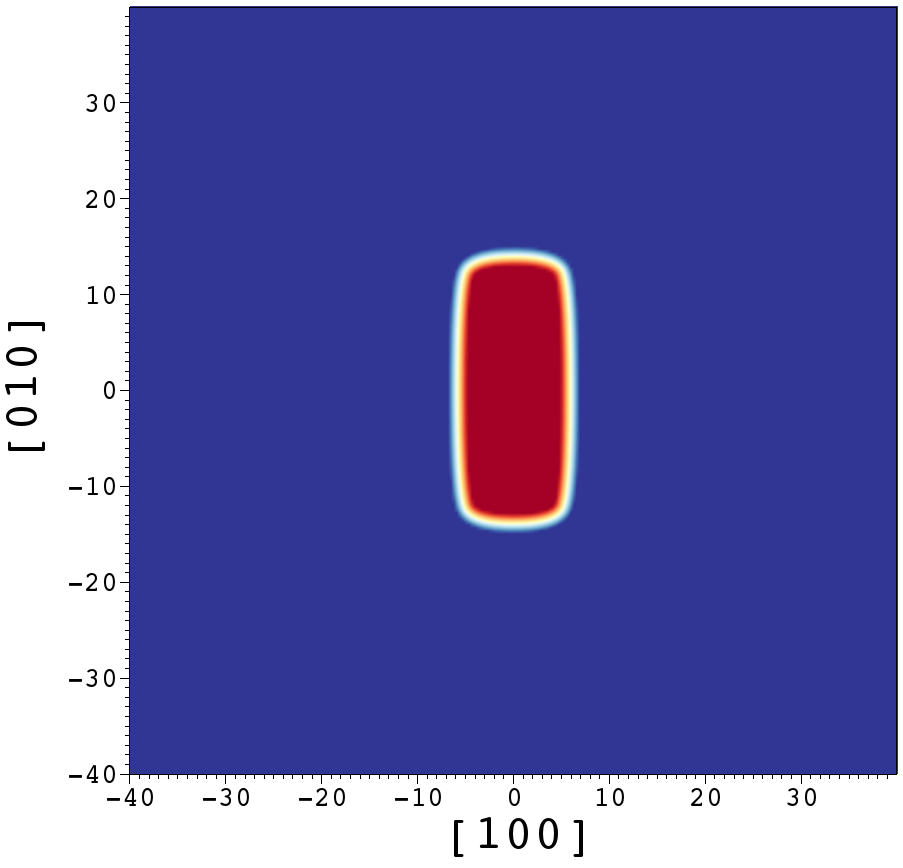}
	\captionof{subfigure}{DNS-ML, (001) plane}
\end{minipage}
\begin{minipage}[t]{0.4\textwidth}
        \centering
	\includegraphics[width=0.9\textwidth]{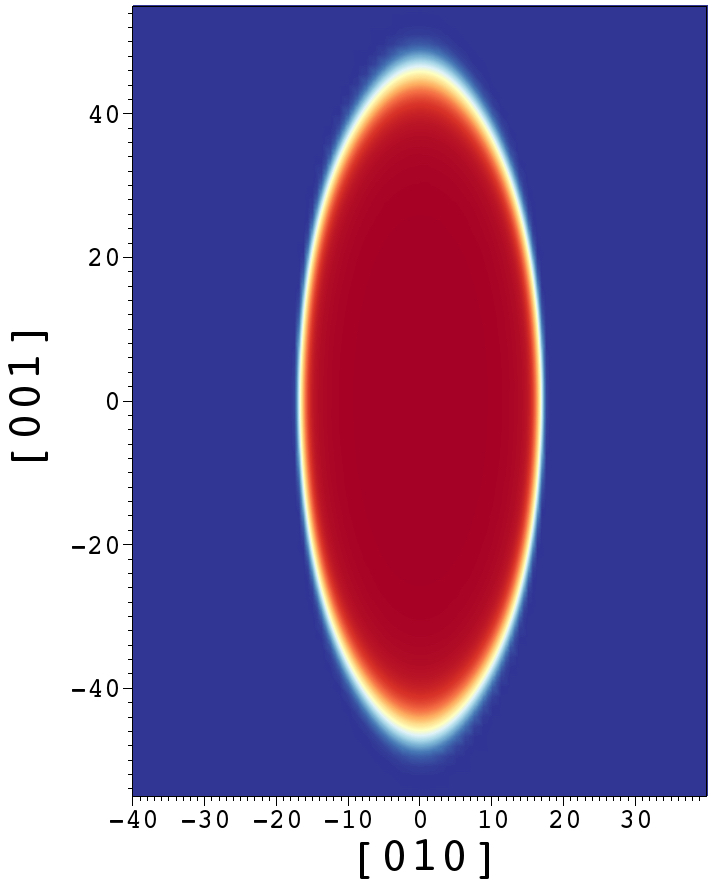}
	\captionof{subfigure}{Phase field, (100) plane}
\end{minipage}
\begin{minipage}[t]{0.4\textwidth}
        \centering
	\includegraphics[width=0.9\textwidth]{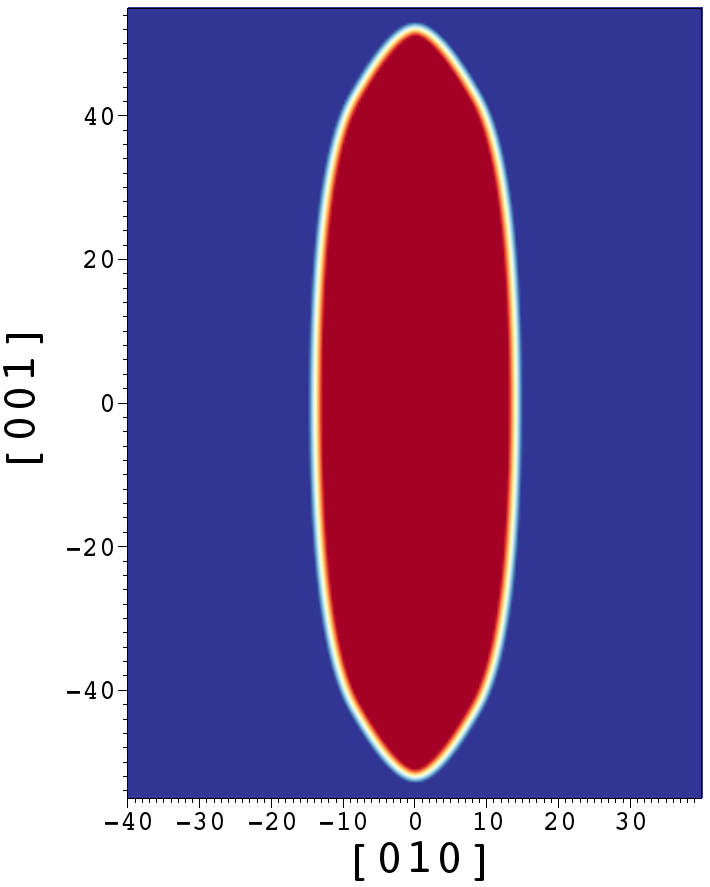}
	\captionof{subfigure}{DNS-ML, (100) plane}
\end{minipage}
        \caption{Comparison of the phase field and DNS-ML methods using 2D slices of the simulation. The precipitate is colored red and the solid-solution is blue.}
	\label{fig:prec_comp}
\end{figure}

\begin{table}[tb]
    \centering
    \caption{Results and computation time for the DNS-ML method and phase field method.}
    \begin{tabular}{c | c c }
    & DNS-ML & PF\\
    \hline
    $v_p$ (nm$^3$) & 23100 & 21500\\
    $c_p$ & 0.12507 & 0.12508\\
    $a$ & 5.87 & 5.93\\
    $b$ & 13.9 & 16.7\\
    $c$ & 52.0 & 45.7\\
    energy (aJ) & -32.1 & -34.0\\
    \hline
    Walltime (hr) & 16.8 & 143\\
    Iterations/steps & 12 & 600\\
    Processes/job & 5 & 240\\
    Max. concurrent jobs & 200 & 1\\
    Approx. total FLOP count & $7\times10^{15}$  & $2\times10^{16}$ 
    \end{tabular}
    \label{tab:results_cv}
\end{table}

The DNS-ML method was implemented using KBNNs. To show the convergence of the DNS-ML method, the lowest high-fidelity DNS energy computation at each iteration is plotted against the iteration index in Figure \ref{fig:convergence_cv}. The method was considered sufficiently converged after twelve iterations, with a computation time of 16.8 hours. The predicted equilibrium values for the features after twelve iterations of the DNS-ML method are presented in Table \ref{tab:results_cv}, along with computational time and resources used. The predicted equilibrium shapes at various iterations of the DNS-ML method are shown in Figure \ref{fig:prec_DNN}. Again, the predicted equilibrium values and shapes are given by the lowest energy high-fidelity DNS computation at each iteration.

\begin{figure}
        \centering
\begin{minipage}[t]{\textwidth}
        \centering
	\includegraphics[width=0.85\textwidth]{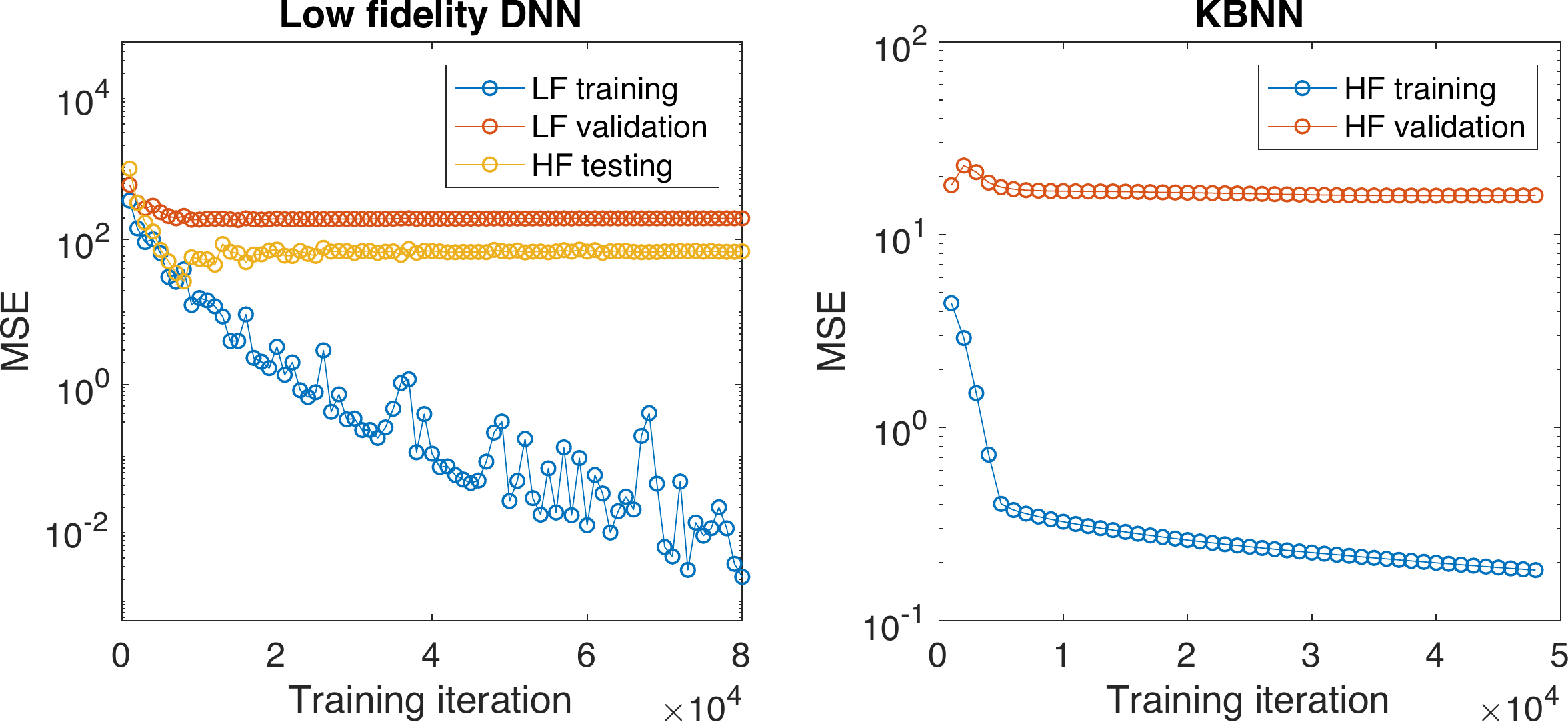}
	\captionof{subfigure}{Data from DNS-ML iteration 1}
\end{minipage}
\begin{minipage}[t]{\textwidth}
        \centering
	\includegraphics[width=0.85\textwidth]{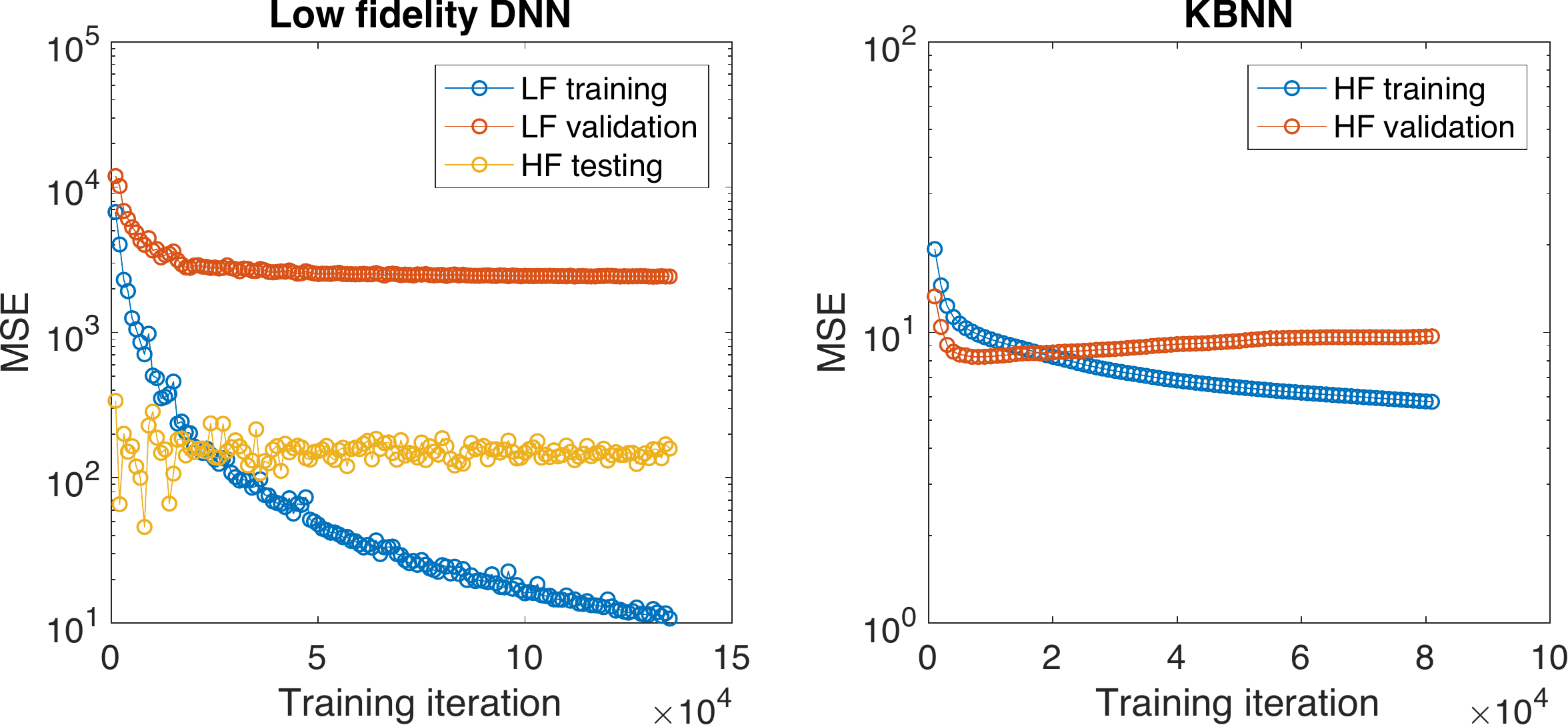}
	\captionof{subfigure}{Data from DNS-ML iteration 12}
\end{minipage}
        \caption{\textcolor{black}{Representative learning curves showing training and validation error are plotted, using data from (a) the first and (b) twelfth iterations of the DNS-ML algorithm for the single precipitate problem. Plots on the left show the training of the standard DNNs to low fidelity data, as well as the test error using high fidelity data. Plots on the right show the training of the KBNNs to the high fidelity data.}}
	\label{fig:learning_curves}
\end{figure}

The FLOP count in Table \ref{tab:results_cv} was estimated using the performance counter tool \texttt{perf} with the event \texttt{pm\_flop} on the Power8 Minsky nodes in the ConFlux cluster. The count was performed for a low-fidelity DNS, a high-fidelity DNS, one time step of the phase field model, and one round of KBNN training and testing. These values were then used to estimate the total FLOP count. Although the phase field model was run on XSEDE Comet and the TensorFlow training and testing of the KBNN on the ConFlux GPUs, the FLOP count for these computations was done on the Power8 nodes to ensure a consistent count method across all simulations. The FLOP count for KBNN training and testing was on the order of $1\times10^{14}$, having a minimal effect on the total count. The equilibrium shapes predicted by the phase field and DNS-ML methods are compared in Figure \ref{fig:prec_comp}.

\textcolor{black}{The learning curves for training and cross-validation appear in Figure \ref{fig:learning_curves}. We note that the plots on the left also include the mean square error when testing the low fidelity DNN on the high fidelity data.  Interestingly, the high fidelity testing error is lower than the low fidelity cross-validation error for both low fidelity DNN learning curves. This is likely due to the high fidelity data being centered more closely around the minimum, where values of the energy and its gradient are low. Consequently, fluctuations may also be expected to be small and the data are more easily trained to. The low fidelity data used in training, on the other hand, fill a much larger region in the parameter space, and values of the energy and its gradient on the periphery can be orders of magnitude higher than the high fidelity data clustered near the minimum. Correspondingly, fluctuations in the low fidelity data are likely to be larger and present difficulties for training, resulting in larger errors far from the minimum. In any case, the improvement gained with the full KBNN over the low-fidelity DNN is apparent for the first as well as the twelfth DNS-ML iteration.
At both initial and late stages of the workflow, the cross-validation error of the KBNN is roughly an order of magnitude lower than the mean squared error from testing the low fidelity DNN on the high fidelity data.
Also notable is that, in the twelfth DNS-ML iteration the ratio of cross-validation to training errors is significantly lower than in the first iteration. This and other differences in the magnitude of the error between DNS-ML iterations can be attributed to the increasing size of the data set and different hyperparameters, as well as the sensitivity analysis that progressively tightens the bounds that define each set $\mathcal{D}_k$.}

\begin{figure}[tb]
        \centering
        \includegraphics[width=0.8\textwidth]{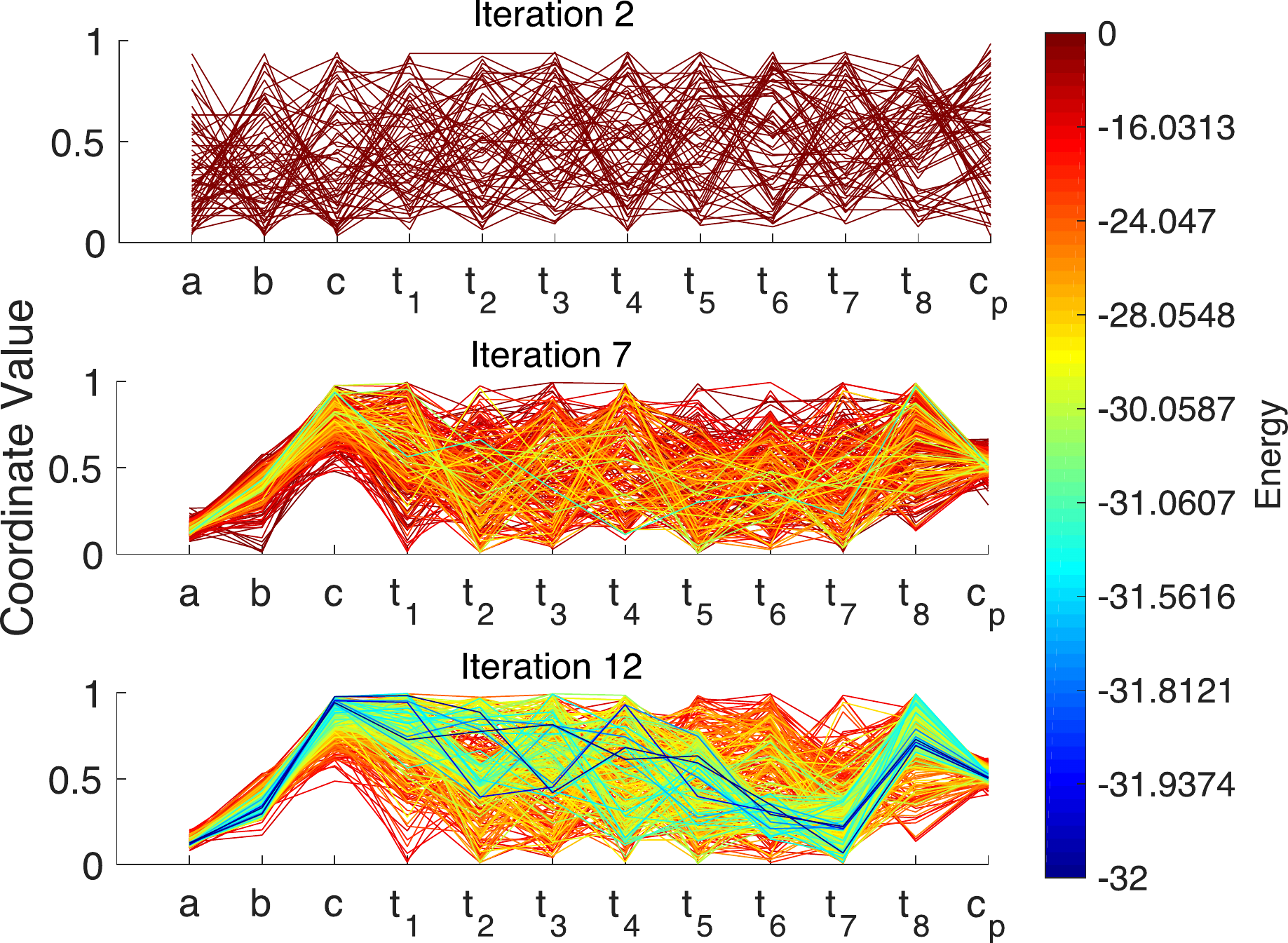}
        	\caption{A parallel coordinates plot of the high-fidelity DNS data used at various iterations of the DNS-ML method. \textcolor{black}{The labels along the $x$-axis are the twelve machine learning features describing the precipitate.} The lines are colored on a log scale according to the energy, with blue representing the lowest energy computations and red the highest energy. As the method converges, lower energy computations are discovered.}
	\label{fig:parallelCoords}
\end{figure}

The high-fidelity DNS computations used at three of the iterations during the DNS-ML algorithm are plotted using parallel coordinates in Figure \ref{fig:parallelCoords}, where each computation has been colored on a log scale according to the total energy on a log scale. The dark blue lines indicate the location of a well in the energy surface. These plots show how the search space focused in on certain features that were more significant than the others, particularly the features $a$, $b$, and $c$ which describe the overall dimensions, and  the precipitate composition, $c_\mathrm{p}$. 
\textcolor{black}{The parameters $t_4$ and $t_5$ appear to have the smallest effect on the total energy. Since each free parameter represents an additional dimension of the search space, it would be worthwhile to remove parameters that show little impact on the quantity of interest (total energy). A possible modification to Algorithm \ref{algo:surrogateopt2} would be to fix shape parameters that maintain a low sensitivity index over multiple DNS-ML iterations. This would gradually reduce the dimension of the search space in an informed manner, thus increasing the speed of convergence and reducing the required number of simulations.}

In comparing the predicted shapes in Table \ref{tab:results_cv} and in Figure \ref{fig:prec_comp}, the overall shape and size are seen to be similar. The cut through (001) plane is quasi-rectangular, while the (100) plane shows an elongated precipitate with slightly peaked tips. The relation $c>b>a$ holds for both predictions. The predicted precipitate compositions are nearly identical. Also, although differences in boundary conditions prevent a direct comparison, the experimental observations in Figure \ref{fig:MgY_STEM} show precipitate shapes that align better with the DNS-ML than the phase field results. We call attention in particular, to the flattening of the shape perpendicular to the [001] direction, which is equivalent to the [0001] direction in Mg. \textcolor{black}{Additional work making a statistical comparison of the DNS-ML and phase field results, with appropriate boundary conditions, to experimental observations would be of interest.}

There is some discrepancy between the two methods in the predicted size and energy. For example, the DNS-ML method predicted a width, $2b$, about a factor of $0.83\times$ and a length, $2c$, about $1.15\times$ of the phase field precipitate. This difference in predictions is largely due to the way interfacial energy is included in the two methods. The DNS-ML method directly takes the given anisotropic interfacial energies and computes the total interfacial energy using a surface integral, as in Equation (\ref{eqn:DNS_IE}). The phase field method first relates the given interfacial energies to the gradient by using the 1D equilibrium solution (Section \ref{sec:gradientenergy}) where elasticity has been neglected. The contribution of the interfacial energy is also divided between the Landau free energy and the gradient free energy. The total interfacial energy is then calculated by integrating the Landau and gradient energies over the domain (see Equations (\ref{eqn:Landau}), (\ref{eqn:grad_energy})-(\ref{eqn:lambda})). 
\textcolor{black}{Furthermore, the varying interface thickness in the phase field results likely also affects the strain and bulk chemical free energies, which involve integrating energy densities that are dependent on the value of the order parameter in the diffuse interface.}
The difference in final energy is also affected by the mesh refinement and coarsening that introduces occasional fluctuations into the total energy over time.

\textcolor{black}{To further explore the difference in the minimum energy found by the two methods, the energy for the equilibrium phase field shape was also computed using the DNS-ML energy formulation from Equation (\ref{eqn:DNS_energy}). The total energy was -31.6 aJ, which is 0.5 aJ higher than the minimum energy of the shape predicted by DNS-ML. The differences in the individual components of the energy were much higher (see Table \ref{tab:pf_diffEnergy}).}

\begin{table}[tb]
    \centering
    \textcolor{black}{\caption{\textcolor{black}{Predicted energies computed for the equilibrium phase field shape, using the phase field energy description in Equation (\ref{eqn:pf_energy}) and the DNS-ML energy description in Equation (\ref{eqn:DNS_energy}).}}
    \begin{tabular}{|c|c c c|c|}
    \hline
        Energy formulation & Bulk chemical & Strain & Interfacial & Total\\
        \hline
        Phase field & -489.3 & 295.2 & 160.1 & -34.0 \\
        DNS-ML & -485.3 & 316.2 & 137.5 & -31.6\\
        \hline
    \end{tabular}
    \label{tab:pf_diffEnergy}}
\end{table}

The above differences notwithstanding, the goal of reaching an equilibrium prediction more rapidly by using the DNS-ML algorithm was clearly met. The required walltime was nearly an order of magnitude less than the phase field simulation. The potential speed up is even greater. Due to limited resources, only 200 or fewer DNS were allowed to run concurrently on ConFlux. Additional resources would allow more DNS to run at the same time, as well as faster computation times for the high fidelity DNS. This would result in a further decrease in walltime. Additionally, the DNS-ML algorithm also resulted in a lower total FLOP count, with the phase field simulation performing over twice as many FLOPs as the DNS-ML algorithm.

\subsection{Two precipitates}
The study can be expanded to multiple precipitates by adding the appropriate features. As an example, we considered two symmetric precipitates of equal size and shape.
\textcolor{black}{However, a more complete treatment could allow for asymmetric precipitates of unequal size and, thus, would include additional features.}
In addition to the shape and composition features used in the single precipitate problem, we also included three features defining the relative position of the two precipitates: distance, $\rho$; azimuthal angle relative to [100], $\theta$; and polar angle relative to [001], $\phi$. Due to symmetry, we took $\theta\in[0,\pi/2]$ and $\phi\in[0,\pi/2]$.

\begin{figure}[tb]
        \centering
\begin{minipage}[t]{0.24\textwidth}
        \centering
	\includegraphics[width=0.99\textwidth]{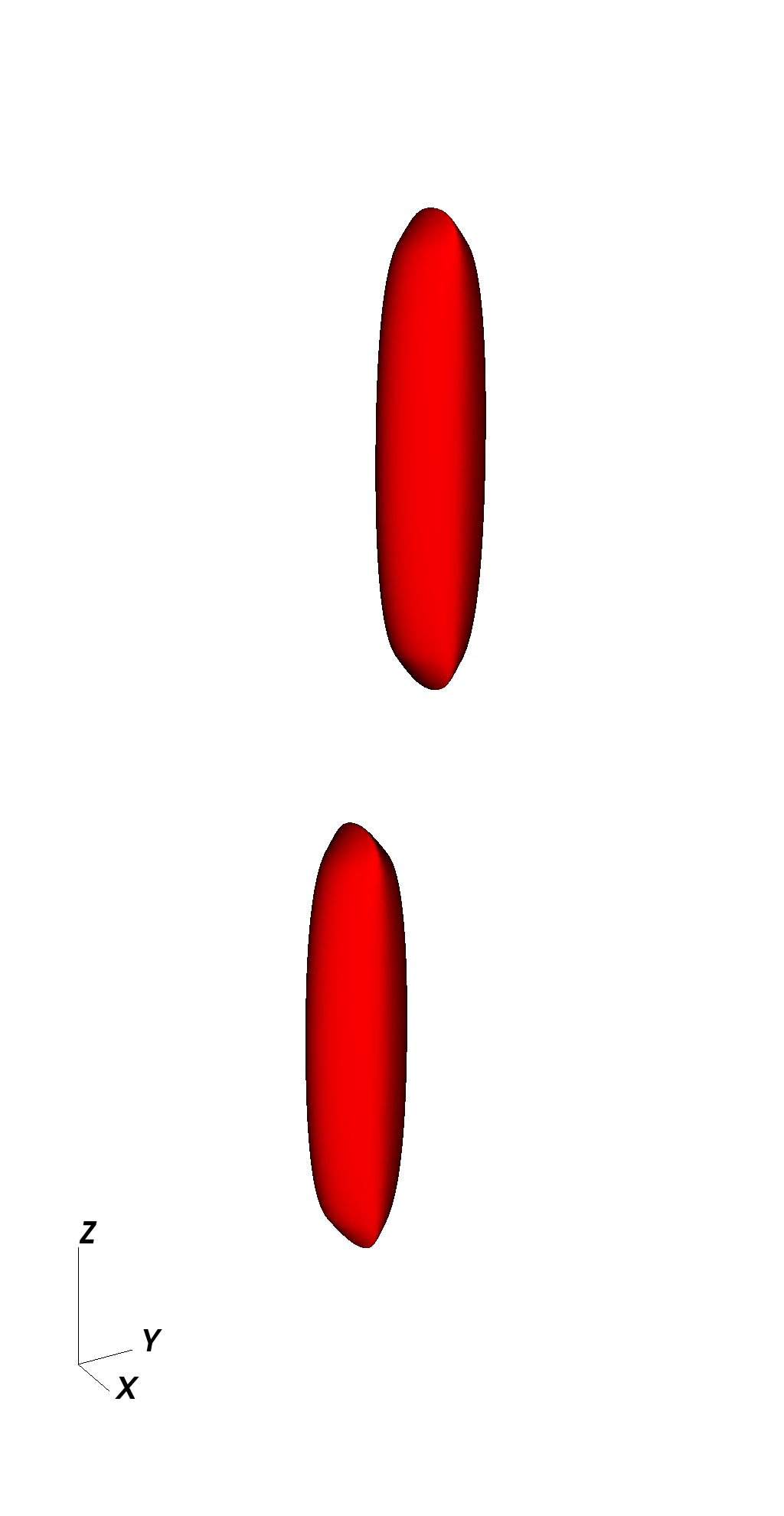}
	\captionof{subfigure}{Iteration 1}
\end{minipage}
\begin{minipage}[t]{0.24\textwidth}
        \centering
	\includegraphics[width=0.99\textwidth]{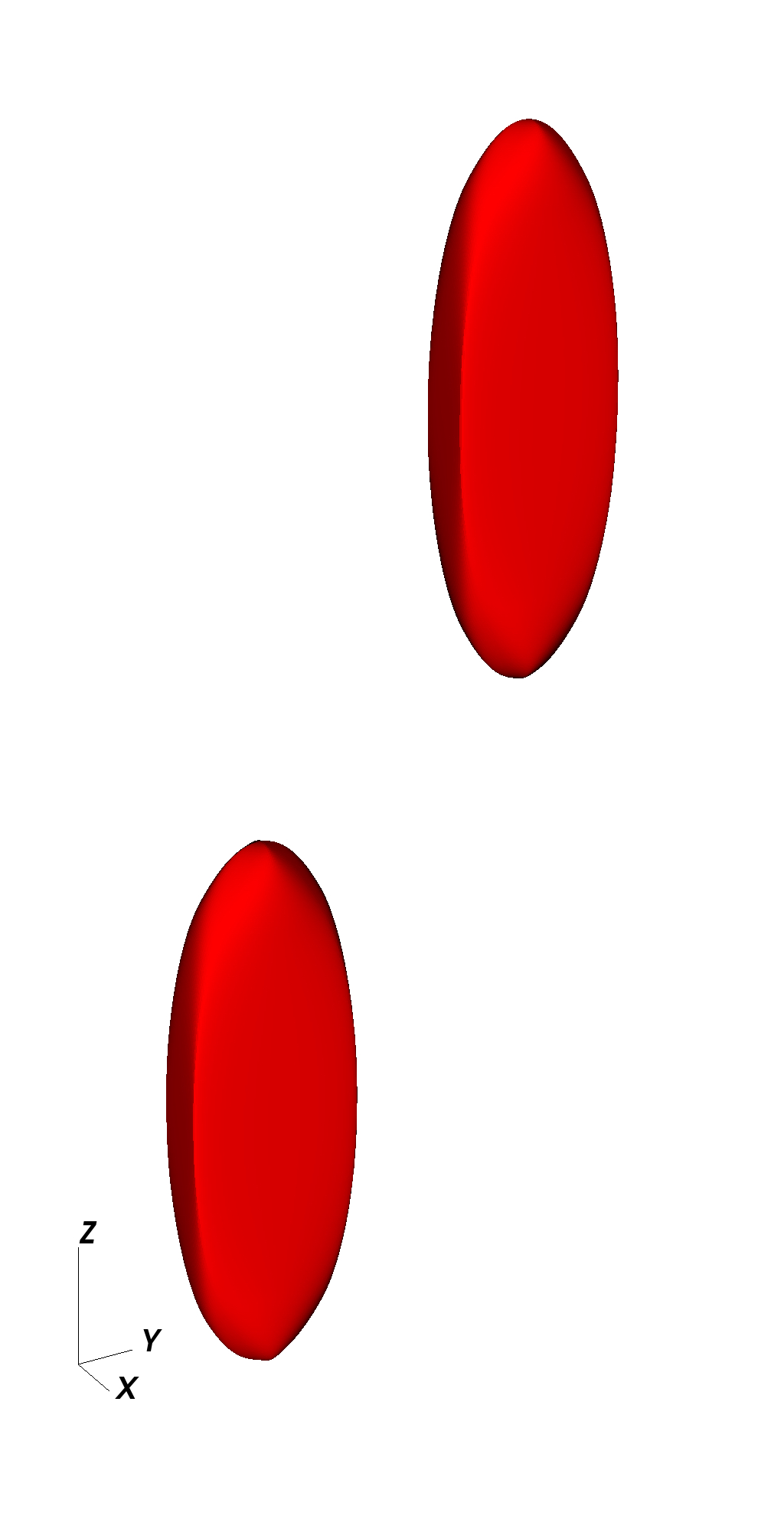}
	\captionof{subfigure}{Iteration 7}
\end{minipage}
\begin{minipage}[t]{0.24\textwidth}
        \centering
	\includegraphics[width=0.99\textwidth]{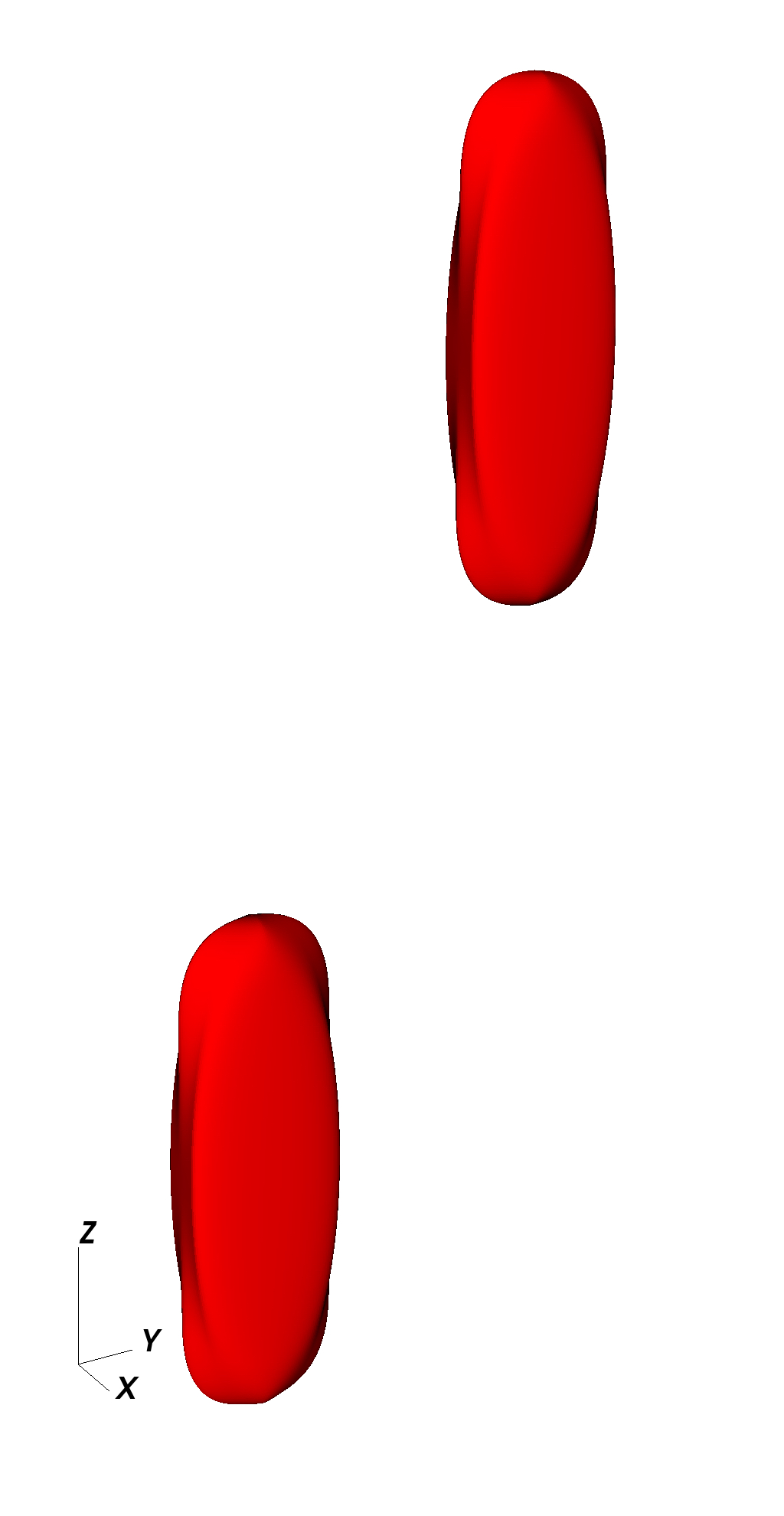}
	\captionof{subfigure}{Iteration 14}
\end{minipage}
\begin{minipage}[t]{0.24\textwidth}
        \centering
	\includegraphics[width=0.99\textwidth]{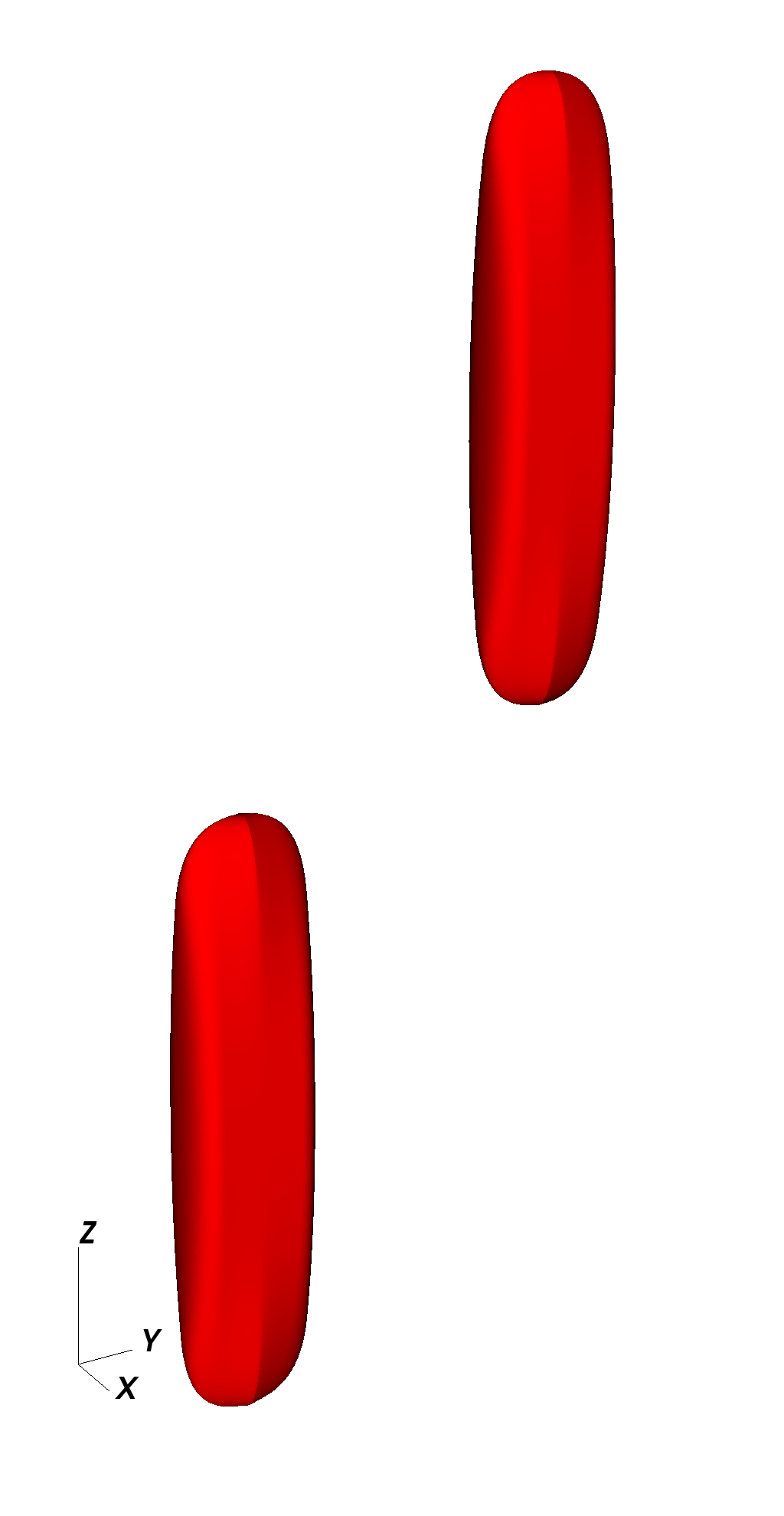}
	\captionof{subfigure}{Iteration 21}
\end{minipage}
        \caption{The predicted precipitate shapes at various iterations of the DNS-ML method with two precipitates.}
	\label{fig:prec_DNN2}
\end{figure}

We considered only precipitate pairs that do not intersect and do not extend beyond the physical domain of $[-40,40]\times[-40,40]\times[-110,110]$. These constraints were checked before completing the DNS for each potential set of feature values taken from the Sobol$^\prime$ sequence. The multifidelity model approximating the energy surface, as in the single precipitate problem, was created during each iteration of the DNS-ML algorithm. The minimum of the approximate energy surface was found using a constrained optimization algorithm. Three types of inequality constraints are applied: 1) each feature is bounded by the current domain of interest, 2) the bounding box containing both precipitates must remain within the physical domain, and 3) the precipitates must not intersect. The third constraint on intersection is difficult to impose exactly due to the parametric surface representation of the precipitate shapes. It was simplified by instead preventing the intersection of two ellipsoids with the same location and dimensions as the B-spline surfaces. The constrained optimization was done using TensorFlow's ScipyOptimizerInterface, which leverages the SLSQP SciPy optimizer.  The sequence of evolving shapes for the two precipitate problem by the DNS-ML method appear in Figure \ref{fig:prec_DNN2}.

\begin{figure}[tb]
        \centering
\begin{minipage}[t]{0.48\textwidth}
        \centering
	\includegraphics[width=0.9\textwidth]{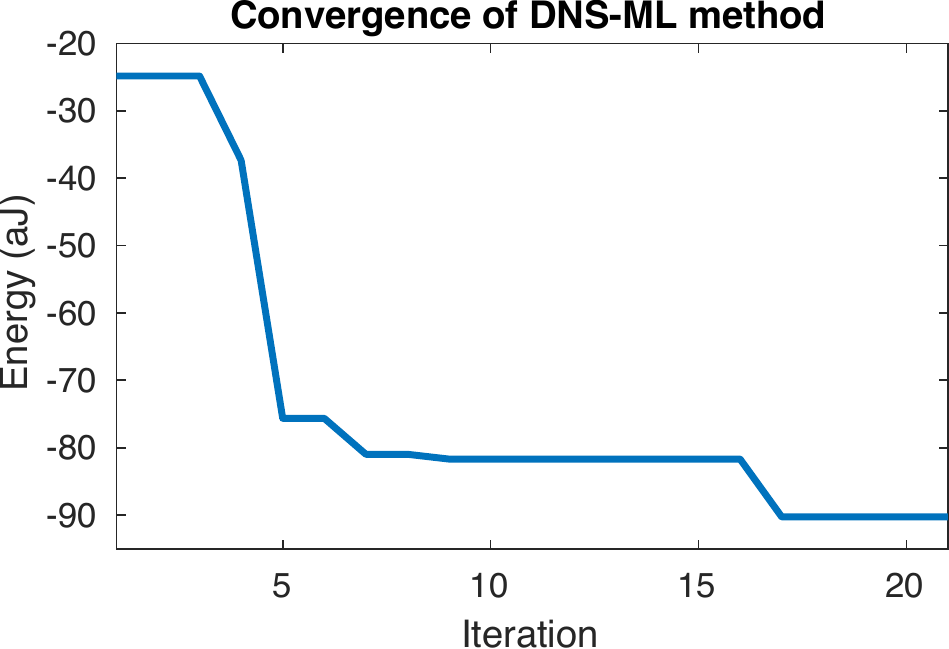}
	\captionof{subfigure}{Two precipitates}
\end{minipage}
\begin{minipage}[t]{0.48\textwidth}
        \centering
	\includegraphics[width=0.9\textwidth]{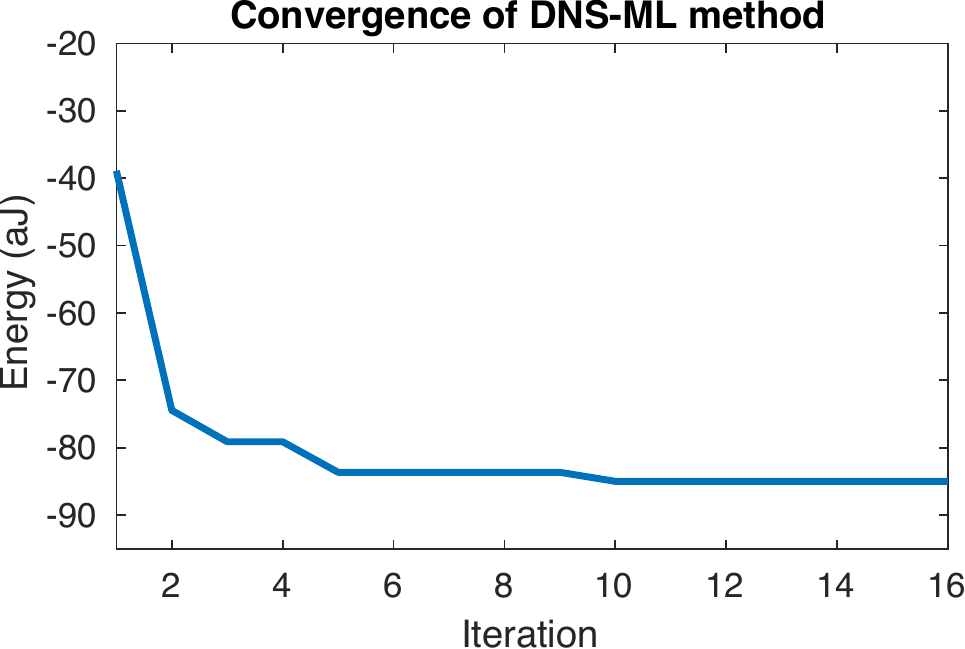}
	\captionof{subfigure}{Single precipitate}
\end{minipage}
        \caption{Convergence of the DNS-ML method for precipitates in the physical domain of $[-40,40]\times[-40,40]\times[-110,110]$.}
	\label{fig:convergence_cv2}
\end{figure}

It is possible for a lower energy to be achieved when the two precipitates merge. To check this possibility, we again ran the DNS-ML method for the single precipitate case, but with the enlarged physical domain used in the two precipitate problem. The results are compared in Table \ref{tab:results_cv2}. We found that the lowest energies were similar, but slightly lower in the two precipitate simulation. This leads to the conclusion that it is energetically favorable to remain as two separate precipitates. The convergence of the double and singe precipitates are shown in Figure \ref{fig:convergence_cv2}.

\begin{table}[tb]
    \centering
    \caption{Results and computation time for the DNS-ML method and phase field method for the two iprecipitate problem.}
    \begin{tabular}{c | c c c }
    & DNS-ML & DNS-ML & PF\\
    & (two prec.) & (one prec.) & \\
    \hline
    total volume (nm$^3$) & 50400 & 44900 & 43700\\
    $c_p$ & 0.1249 & 0.1252 & 0.1251\\
    average $a$ & 7.70 & 6.05 & 5.8\\
    average $b$ & 11.4 & 18.1 & 16\\
    average $c$ & 51.6 & 73.1 & 44\\
    $\rho$ & 127 & N/A & 120\\
    $\theta$ & 1.57 & N/A & 1.2\\
    $\phi$ & 0.45 & N/A & 0.03\\
    energy (aJ) & -90.3 & -85.0 & -119\\
    \hline
    Walltime (hr) & 78.1 & 54.6 & 224\\
    Iterations/steps & 21 & 16 & 890\\
    Max. processes/job & 10 & 10 & 480\\
    Max. concurrent jobs & 200 & 200 & 1\\
    Approx. total FLOP count & $4\times10^{16}$ & $2\times10^{16}$ & $1\times10^{17}$ 
    \end{tabular}
    \label{tab:results_cv2}
\end{table}

\begin{figure}[tb]
        \centering
\begin{minipage}[t]{0.48\textwidth}
        \centering
	\includegraphics[width=0.7\textwidth]{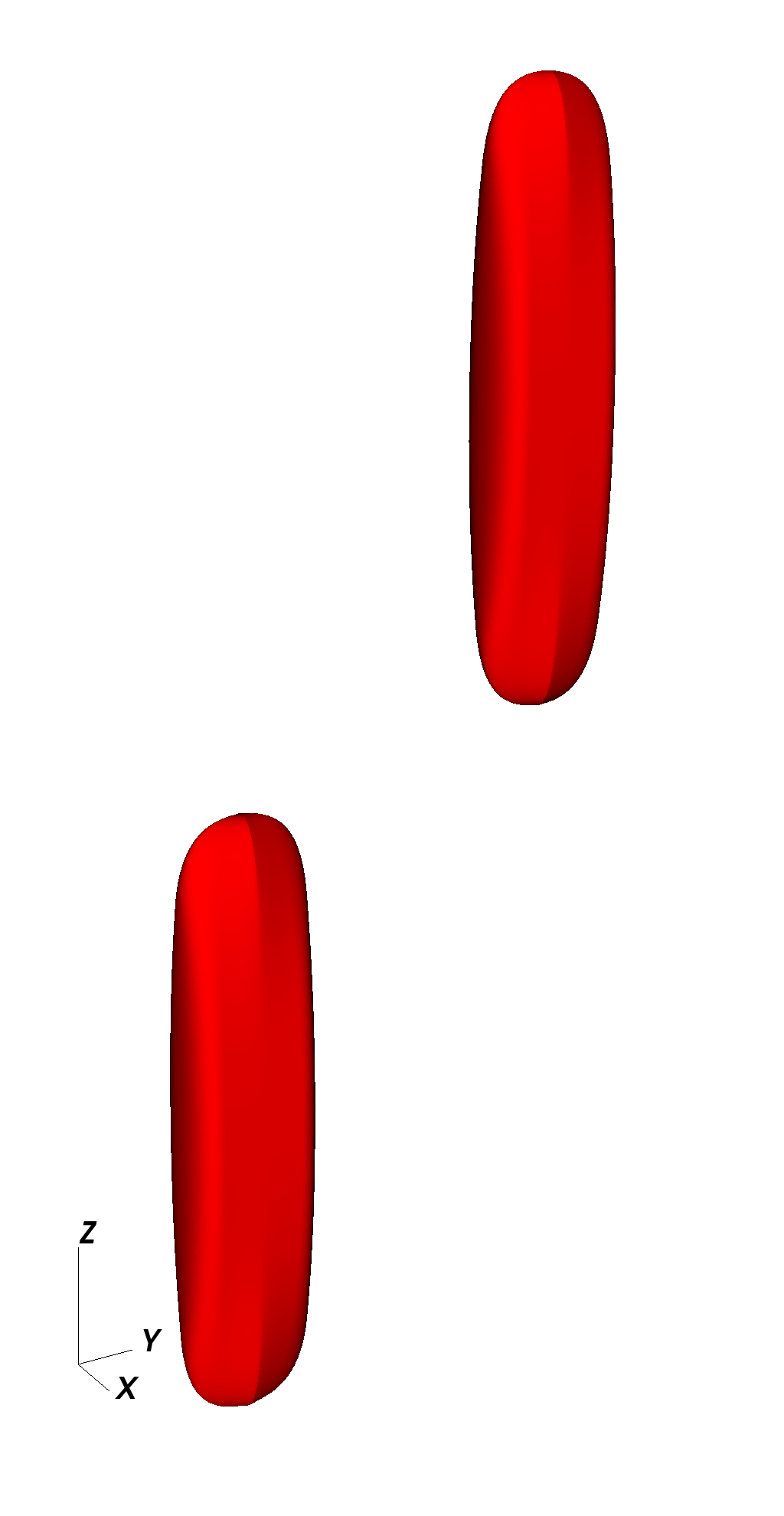}
	\captionof{subfigure}{DNS-ML}
\end{minipage}
\begin{minipage}[t]{0.48\textwidth}
        \centering
	\includegraphics[width=0.7\textwidth]{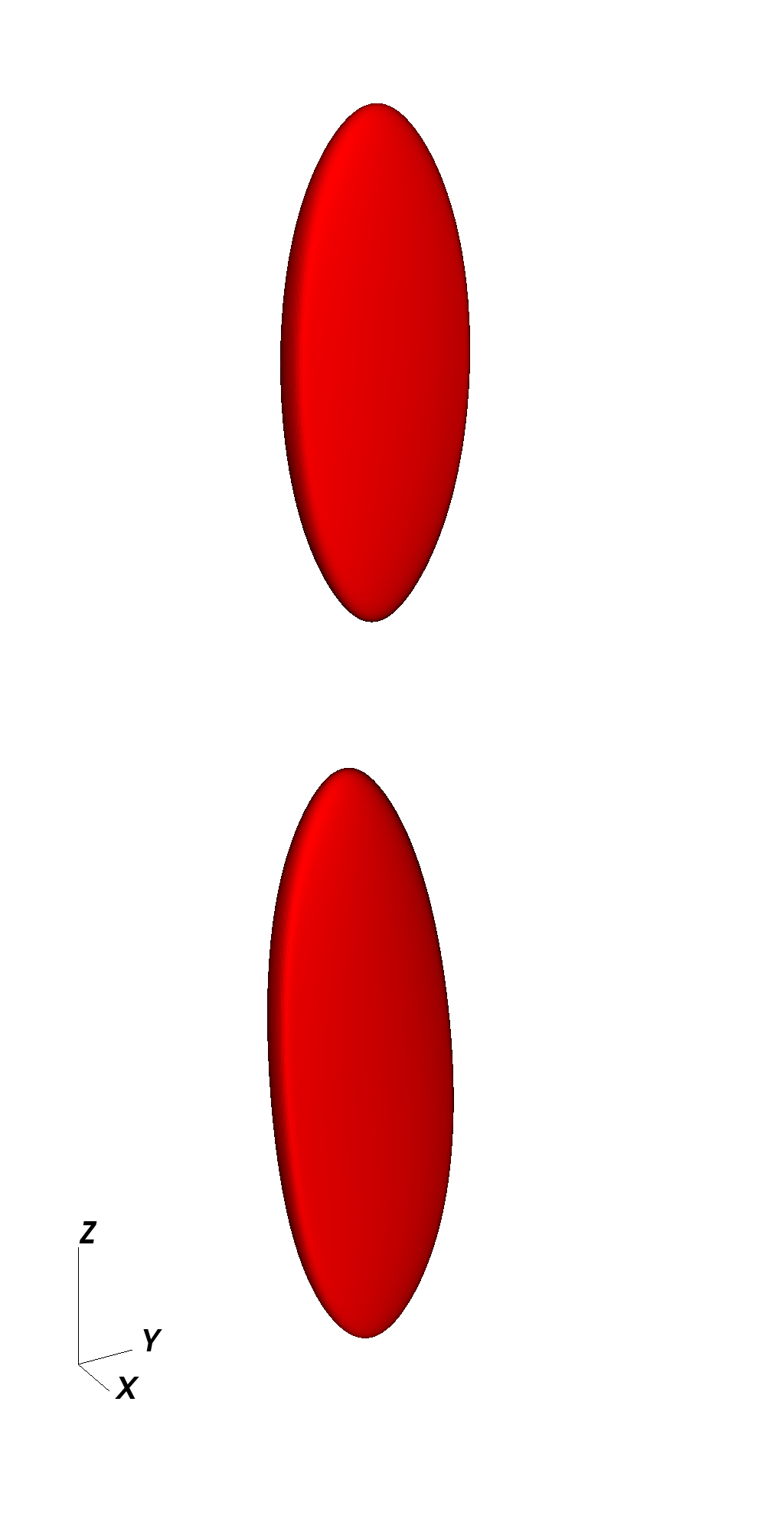}
	\captionof{subfigure}{Phase field}
\end{minipage}
        \caption{Comparison of the predicted shapes and positions by the DNS-ML and phase field methods with two precipitates.}
	\label{fig:comparisonTwoPrec}
\end{figure}

Wider differences are seen between the DNS-ML and phase field results in the two precipitate problem than in the single precipitate simluations. Both methods predict that the two precipitates lie in roughly the same (100) plane, but DNS-ML shows an offset along the [010] direction that is not seen in phase field (see Figure \ref{fig:comparisonTwoPrec}). It is possible that the energy landscape has local minima corresponding to each of these two configurations, and the two methods converged to different energy wells. Much of the difference in the energies reported in Table \ref{tab:results_cv2} is likely due to the slight increase in average composition in the phase field due to mesh refinement and coarsening.
\textcolor{black}{It is noted that the phase field simulation again reports a lower energy than the result found with DNS-ML. Calculating the total energy based on the location and average phase field shape, but with the DNS-ML energy functional, resulted in a total energy of -62.2 aJ, nearly 30 aJ higher than the DNS-ML minimum energy for the two precipitate problem.}
We also note that the DNS-ML results are in better agreement with the experimental images in Figure \ref{fig:MgY_STEM}, as also noted in connection with Figure \ref{fig:prec_comp}. However, we have not completed a statistically rigorous comparison.

We again see the relation $a<b<c$ in the results from both DNS-ML and phase field. The phase field method shows a consistent shape when comparing the single and multiple precipitate computations. The DNS-ML method with two precipitates does not result in the peaked ends seen in the single precipitate results, but it does maintain the quasi-rectangular shape in the (001) plane.

While the required walltime is clearly less for the DNS-ML method in comparison with the phase field method, the speed up of about $3\times$ is less than was observed for the single precipitate problem. This is partly due to the increased computational cost to construct the signed-distance function based on two, instead of one, parametric surface in the DNS.
\textcolor{black}{When including the DNS-ML computation checking for a single, merged precipitate, the speed up over phase field reduces to $1.7\times$. Furthermore, it is not clear beforehand how many precipitates would exist at equilibrium. It is possible that three or four would have a lower energy than two precipitates. Therefore, when considering the case of multiple precipitates, it may be necessary to incorporate an optimization method that allows for the optimization of discrete values, such as genetic algorithms or simulated annealing.}

Also, as observed in Figure \ref{fig:convergence_cv2}, the convergence is less consistent than the single precipitate problem, with several iterations resulting in no decrease in the predicted minimum energy. This could be due to an inadequately trained multifidelity model during certain iterations, influenced by the higher number of features in the two precipitate problem, or simply a more complex energy landscape. The FLOP count for the DNS-ML computations, however, remained less than half the number of FLOPs performed by the phase field simulation.

\section{Conclusions}
\label{sec:conclusions}

In this preliminary communication we have explored the feasibility of using machine learning techniques to detect equilibrium states in a physical system. For the precipitate morphology problem, this translates to finding minima of a free energy landscape. The free energy induces a form of surrogate optimization in this implementation. We have used a multifidelity model based on Deep Neural Networks as the surrogate model for the free energy in the optimization routine.

The DNS-ML algorithm, when used for a single precipitate, compares favorably with the phase field method commonly used in the computational study of precipitate morphology. The predicted shapes and compositions of the DNS-ML method were similar to the near-equilibrium results found by phase field. Furthermore, the DNS-ML algorithm required roughly an order of magnitude less computation time than phase field. The phase field dynamics provide information on the precipitate's configurations as it grows and evolves with time. The DNS-ML method also explores these configurations, even if not in a time-sequential manner. With total free energies also being computed, the configurations can be ordered in the direction of declining free energy as we have done. The large number of configurations explored by the DNS-ML method provides an overview of the entire free energy landscape, rather than only the path taken by a solution, as is the case with the phase field method.

Within the DNS-ML method, machine learning provides a distinct contribution in two areas. The first is in performing the sensitivity analysis. The Monte Carlo algorithm used in the variance-based sensitivity analysis requires tens of thousands of data points to converge. This number of data points is feasible when the low-fidelity model is a function evaluation that reasonably captures important trends. However, the computation time required can become intractable even with the low-fidelity model when, as in this work, the low-fidelity model is simply a coarse-meshed variant of the high-fidelity model. The machine learned model bypasses this difficulty by providing functions in the form of DNNs that can rapidly be evaluated on the thousands of required points.

The second major advantage gained through machine learning appears in the multifidelity model. It is possible to use, for example, a standard DNN as the surrogate model in the algorithm presented in this work. In such a case, high-fidelity data would be required for all training data. However, in our experience \textcolor{black}{using this simpler model in earlier versions of the DNS-ML,} the standard DNN cannot predict a minimum that is significantly better than the lowest energy high-fidelity data point. The large number of points needed to train an accurate surface makes the surface somewhat irrelevant, at least in the case where the only goal is to find a minimum. The relevance of the surface returns, however, when it can be accurately constructed with only sparse high-fidelity data. Machine learning allows us to learn the relation between the low-fidelity and high-fidelity data, thus requiring relatively few high-fidelity data points. By rapidly constructing a multifidelity model that approximates the high-fidelity data, we can predict a minimum that is more accurate than the minimum energy point of the sparse high-fidelity data set.

This work will serve as a seed for explorations of machine learning to predict the equilibrium morphology of single precipitates for a range of matrix and precipitate crystal structures, elasticities and interfacial energies.

\section{Acknowledgements}
\label{sec:acknowledgements}
We thank Emmanuelle Marquis and Ellen Solomon for their insight and the contribution of the STEM images presented here. We also thank Anirudh Natarajan for the calculation of the $\beta'$ precipitate elasticity constants. We gratefully acknowledge the support of Toyota Research Institute, Award \#849910, ``Computational framework for data-driven, predictive, multi-scale and multi-physics modeling of battery materials." This work has also been supported in part by NSF DMREF grant: DMR1436154, ``DMREF: Integrated Computational Framework for Designing Dynamically Controlled Alloy-Oxide Heterostructures." Support has also been provided by the U.S. Department of Energy, Office of Basic Energy Sciences, Division of Materials Sciences and Engineering under Award \#DE-SC0008637 that funds the PRedictive Integrated Structural Materials Science (PRISMS) Center at University of Michigan. Computing resources were provided in part by the NSF via grant 1531752 MRI: Acquisition of Conflux, A Novel Platform for Data-Driven Computational Physics (Tech. Monitor: Ed Walker). This work also used the Extreme Science and Engineering Discovery Environment (XSEDE) Comet at the San Diego Supercomputer Center through allocation TG-MSS160003.

\bibliographystyle{unsrt}
\bibliography{references}

\end{document}